\documentclass{nature}

\usepackage{graphicx}
\usepackage{multibib}
\usepackage{filecontents}
\usepackage{placeins}

\newcites{mp}{References}
\newcites{sm}{Additional References}

\title{Tidal Evolution of the Moon from a High-Obliquity High-Angular-Momentum Earth}

\author{Matija \' Cuk$^1$, Douglas P. Hamilton$^2$, Simon J. Lock$^3$ and Sarah T. Stewart$^4$}

\begin{document}

\maketitle

\begin{affiliations}
 \item Carl Sagan Center, SETI Institute, 189 North Bernardo Avenue, Mountain View, CA 94043, USA
 \item Department of Astronomy, University of Maryland, Physical Sciences Complex, College Park, MD 20742, USA
 \item Department of Earth and Planetary Sciences, Harvard University, 20 Oxford Street, Cambridge, MA 02138, USA
 \item Department of Earth and Planetary Sciences, University of California Davis, One Shields Avenue, Davis, CA 95616, USA
\end{affiliations}

\begin{abstract}
In the giant impact hypothesis for lunar origin, the Moon accreted from an equatorial circum-terrestrial disk; however the current lunar orbital inclination of 5$^{\circ}$ requires a subsequent dynamical process that is still debated \citemp{Touma:1998, Ward:2000, Pahlevan:2015}. In addition, the giant impact theory has been challenged by the Moon's unexpectedly Earth-like isotopic composition \citemp{Burkhardt:2015, Young:2016}. Here, we show that tidal dissipation due to lunar obliquity was an important effect during the Moon's tidal evolution, and the past lunar inclination must have been very large, defying theoretical explanations. We present a new tidal evolution model starting with the Moon in an equatorial orbit around an initially fast-spinning, high-obliquity Earth, which is a probable outcome of giant impacts. Using numerical modeling, we show that the solar perturbations on the Moon's orbit naturally induce a large lunar inclination and remove angular momentum from the Earth-Moon system. Our tidal evolution model supports recent high-angular momentum giant impact scenarios to explain the Moon's isotopic composition \citemp{Cuk:2012, Canup:2012, Lock:2016} and provides a new pathway to reach Earth's climatically favorable low obliquity. 

\end{abstract}

The leading theory for lunar origin is the giant impact \citemp{Hartmann:1975,Cameron:1976}, which explains the Moon's large relative size and small iron core. Here we refer to the giant impact theory in which the Earth-Moon post-impact angular momentum (AM) was the same as it is now (in agreement with classic lunar tidal evolution studies\cite{Goldreich:1966, Touma:1994}) as ``canonical". In the canonical giant impact model\cite{Canup:2001}, a Mars-mass body obliquely impacts the proto-Earth near the escape velocity to generate a circum-terrestrial debris disk. The angular momentum of the system is set by the impact, and the Moon accretes from the disk, which is predominantly ($>60$~wt\%) composed of impactor material. However, Earth and the Moon share nearly identical isotope ratios for a wide range of elements, and this isotopic signature is distinct from all other extraterrestrial materials \citemp{Burkhardt:2015, Young:2016}. Because isotopic variations arise from multiple processes \cite{Burkhardt:2015}, the Moon must have formed from, or equilibrated with, Earth's mantle \cite{Meier:2012, Young:2016}. Earth-Moon isotopic equilibration in the canonical model has been proposed by Pahlevan and Stevenson \cite{Pahlevan:2007}, but has been questioned by other researchers \cite{Melosh:2014}, who suggest that the large amount of mass exchange required to homogenize isotopes could lead to the collapse of the proto-lunar disk.

{\' C}uk and Stewart \cite{Cuk:2012} proposed a new variant of the giant impact that is based on an initially high AM Earth-Moon system. In this model, a late erosive impact onto a fast-spinning proto-Earth produced a disk that was massive enough to form the Moon, and was composed primarily of material from Earth, potentially satisfying the isotopic observations. Canup\cite{Canup:2012} presented a variation of a high-AM origin in which a slow collision between two similar-mass bodies produces a fast-spinning Earth and disk with Earth-like composition. Subsequently, Lock et al. \cite{Lock:2016} have argued that a range of high-energy, high-AM giant impacts generate a particular post-impact state where the Earth's mantle, atmosphere and disk are not dynamically isolated from each other, enabling widespread mixing and equilibration between the accreting Moon and Earth. After the impact, these high-AM models require a mechanism to remove AM to be consistent with the current Earth-Moon system. {\' C}uk and Stewart\cite{Cuk:2012} originally proposed the excess AM was lost during tidal evolution of the Moon via the evection resonance between Earth's orbital period and precession of the Moon's perigee \cite{Cuk:2012}. More recently, Wisdom and Tian\cite{Wisdom:2015} found that the evection near-resonance can reduce the system's AM to the present value for a wider range of tidal parameters than explored by {\' C}uk and Stewart\cite{Cuk:2012}. However, AM loss through the evection resonance is still confined to a subset of possible tidal parameters for Earth and the Moon, and the high-AM giant impact scenario requires a robust mechanism for reproducing the present-day system. In this work, we propose a more plausible model for lunar tidal evolution that removes AM, but is mainly motivated by another major problem for lunar origin, the Moon's orbital inclination. 

The Moon's orbit is currently inclined by about 5$^{\circ}$, but studies of its tidal evolution \citemp{Goldreich:1966, Touma:1994} have found that the inclination would have to have been at least $12^{\circ}$ at formation, if the inclination was primordial. This is at odds with lunar formation from a flat disk in Earth's equatorial plane, which should produce a Moon with no inclination. Hypotheses that have been proposed to explain the lunar inclination include a sequence of luni-solar resonances \cite{Touma:1998}, resonant interaction with the protolunar disk \citemp{Ward:2000}, and encounters with large planetesimals following lunar formation\cite{Pahlevan:2015}. However, past studies of lunar tidal history \cite{Goldreich:1966, Touma:1994, Cuk:2012} ignored the obliquity tides within the Moon, despite the Moon having very large ``forced" obliquity when it was between 30 and 40 Earth radii ($R_E$) due to the lunar spin axis undergoing the Cassini state transition \cite{Ward:1975, Chyba:1989}. Chen and Nimmo \cite{Chen:2016} found that the lunar obliquity tides (driven by Earth's apparent north-south motion relative to the lunar figure) significantly decrease the Moon's orbital inclination. 

To quantify the effect of obliquity tides, we used a semi-analytical tidal model (see Methods Section 1). While Chen and Nimmo \cite{Chen:2016} considered tides within the lunar magma ocean that rely on excitation of Rossby waves \cite{Tyler:2008}, here we considered only the tidal response of the current, ``cold" Moon. If we additionally assume long-term average tidal dissipation within Earth and current lunar tidal properties, we find that the orbital inclination of the Moon must have been substantially higher before the Cassini state transition, possibly as high as 30$^{\circ}$ (Methods Section 1 and Extended Data, ED, Figure 1). If we were to extrapolate this inclination back to the time of lunar formation close to Earth, the Moon must have formed with an orbit inclined over 50$^{\circ}$ to the equator, which is clearly inconsistent with a giant impact origin and suggests that the inclination was acquired after the Moon formed. In addition, the planetesimal encounter hypothesis\cite{Pahlevan:2015} has difficulty producing both the correct lunar eccentricity and inclination simultaneously when lunar inclination damping by lunar obliquity tides and lunar eccentricity excitation by Earth tides are taken into account (Methods Section 2 and ED Figure 2). 

We show that the tidal evolution of the Moon starting with a high-obliquity, high AM Earth can reproduce the current lunar orbit, including the lunar inclination and the Earth-Moon system AM. For any perturbed orbit, there exists a {\it Laplace plane} around which the orbital plane of the perturbed orbit precesses. For close-in moons of oblate planets like Jupiter and Saturn, the Laplace plane is the equatorial plane of the planet, while for outer irregular satellites of these planets that are strongly perturbed by the Sun, the Laplace plane is their planet's heliocentric orbital plane. For the Moon, the Laplace plane undergoes a transition during lunar tidal evolution when the Moon recedes from the inner region dominated by perturbations from Earth's equatorial bulge to the outer region dominated by solar perturbations. In the inner region, the Laplace plane is close to Earth's equator, and in the outer region, the Laplace plane is close to the plane of Earth's heliocentric orbit (the ecliptic). At the transition between these two regimes, the Laplace plane is intermediate between the equator and the ecliptic. The distance at which the Laplace plane transition occurs is approximately\cite{Nicholson:2008},
\begin{equation}
r_L=\Bigl(2 J_2 {M_E \over M_S} R^2_E a_E^3\Bigr)^{1/5},
\label{laplace}
\end{equation}
where $J_2$ is the oblateness moment of Earth, $M_E$ and $M_S$ are Earth's and solar mass, and $a_E$ is Earth's semimajor axis. For a planet like Earth that is in hydrostatic equilibrium, $J_2$ depends on the rotation rate. Therefore, the critical distance for the Laplace plane transition has been moving inward over the course of lunar tidal evolution as Earth's rotation slows down and Earth becomes more spherical. For an Earth-Moon system with 100-180\% the present AM, the Laplace plane transition happens at $16-22 R_E$.

For small and moderate obliquities of Earth (i.e. angles between the equator and the ecliptic), the Laplace plane transition is smooth and does not produce any excitation of lunar eccentricity or inclination. However, the Laplace plane transition causes orbital instabilities for obliquities above $68.9$ degrees\cite{Tremaine:2009}. Satellites on circular orbits around high-obliquity planets migrating through the Laplace plane transition can acquire substantial eccentricities and inclinations. This excitation is driven by solar secular perturbations that operate at high inclinations (``Kozai resonance"\cite{Kozai:1962}). For high planetary obliquities, satellites close to the Laplace plane transition with low ``free" inclinations (to the local Laplace plane) still experience solar Kozai perturbations, as their orbits have large instantaneous inclinations relative the the ecliptic \cite{Tamayo:2013}. A related mechanism produces complex dynamics previously found by Atobe and Ida \cite{Atobe:2007} for the tidal evolution of hypothetical high-obliquity Earth-like planets with large moons. Atobe and Ida \cite{Atobe:2007} also found that the mass of a Moon-sized satellite has a significant effect on the dynamics of the system, enabling stagnation or even reversal of tidal evolution and large-scale AM loss from the system; however, they used an averaged model that did not track eccentricity and could not capture the full dynamics of the Laplace plane transition. 

In order to study the tidal evolution of the Moon from a high-obliquity Earth followed by inclination damping at the Cassini state transition, we wrote a specialized numerical integrator {\sc r-sistem} which resolves lunar rotation and therefore fully models lunar obliquity tides (Methods Section 3). Figure 1 (see also Supplementary Information Video 1) shows the early tidal evolution for two simulations that assume that the Earth-Moon system initially had 1.8 times its current AM, as proposed by {\' C}uk and Stewart \cite{Cuk:2012}, but with an initial obliquity to the ecliptic of $70^{\circ}$. Solar perturbations induce significant lunar eccentricity when the Moon reaches the Laplace plane transition at about 17~$R_E$, triggering strong eccentricity-damping satellite tides which shrink the semimajor axis and approximately balance the outward push of Earth tides (first 16 Myr of Fig. 1). As eccentric orbits have less AM than circular ones, AM is removed from the lunar orbit and transferred to Earth's heliocentric orbit; Earth tides in turn transfer AM from Earth's spin to the lunar orbit, while satellite tides do not change the AM of the Earth-Moon system. During this prolonged stalling of lunar tidal evolution, the Moon acquires large inclination (over $30^{\circ}$), while the obliquity of Earth decreases. In the later part of this complex period of the lunar orbital history, lunar eccentricity is excited by secular near-resonances between lunar inclination and eccentricity and Earth's oblateness (see Methods Section 4, ED Figures 3-4). Depending on the exact tidal parameters used for Earth and the Moon, Earth's obliquity can reach that required to match the present value  ($<20^{\circ}$)\cite{Rubincam:2016}, while the AM of the system also matches the present value (0.35 in units of $\alpha_E \sqrt{G M_E^3 R_E}$; see Fig. 1 caption for definitions). Figure 2 presents results of the Laplace plane transition simulations using different tidal parameters for Earth and the Moon and a different initial AM of Earth (ED Figures 5-6). Large AM loss, high lunar inclination and low terrestrial obliquity is a common outcome.

In Fig. 3, we explore the early part of the Laplace plane transition for different initial obliquities of Earth (with the same the total AM). Cases with initial obliquities of $65^{\circ}$ and $75^{\circ}$ are similar to Fig. 1, with larger obliquity leading to larger AM loss in the early stages of the Laplace plane transition. The simulation with initial obliquity of $80^{\circ}$ experiences stalling and reversal of tidal evolution, with the Moon eventually falling back onto Earth. This evolution agrees with prior results\cite{Atobe:2007}, and we expect this outcome for all obliquities larger than $80^{\circ}$. We also found that the Moon evolving from Earth with initial obliquity of $60^{\circ}$ does not experience any instability at the Laplace plane transition. As the orientation of terrestrial planets' spin axes is likely determined by giant impacts, their poles should be randomly distributed on a sphere. About a third of all Earth-like planets should have obliquities within $60^{\circ}-80^{\circ}$ or $100^{\circ}-120^{\circ}$, and, if they have large moons, would lose AM and obliquity (without losing the moon) at the Laplace plane transition. Furthermore, Kokubo and Genda\cite{Kokubo:2010} found that planets with large AM are a common outcome of terrestrial planet formation. 

Once the Moon has passed through the Laplace plane transition, it continues to recede from Earth and lunar rotation passes through the Cassini state transition\cite{Ward:1975}. Using {\sc r-sistem}, we find that the Moon likely spent some time in a non-synchronous rotation state when close to the Cassini state transition (Fig. 4, ED Figs. 7-8) and that transitions between the rotation states can be triggered by various resonances or impacts (Methods Section 5, ED Figs. 9-10, and SI Video 2). Regardless of the nature of the lunar rotation state, the Moon's obliquity is very high during the Cassini state transition and immediately following it, leading to damping of lunar inclination. While the implementation of high-obliquity satellite tides in a fully numerical integrator is challenging (Methods Section 3), we find that the lunar inclination damps from 30$^{\circ}$ (obtained during the Laplace plane transition) to its present value if we assume the long-term average tidal properties for Earth and a relatively non-dissipative, solid Moon (Fig. 4). 

The rotational dynamics of the Moon is strongly dependent on the Moon's global shape. The early Moon probably had little strength, and its shape was in equilibrium with tidal forces \cite{GB:2006}. In Fig. 1, we modeled the Moon as a rigid body for numerical tractability, but we periodically reset its figure to match an equilibrium shape for that distance from Earth, assuming synchronous rotation \cite{Keane:2014}. This assumption of a hydrostatic-like shape results in low obliquities in the Cassini state 1 when the Moon is close to Earth \cite{Ward:1975}. Since the current triaxial shape of the Moon matches the order of magnitude of tidal deformation expected at $23-26 R_E$ \cite{GB:2006}, we assumed that the Moon is rigid and has the present-day principal moments beyond $25 R_E$. Our orbital history shown in Fig. 1 may be consistent with the proposal that the current lunar shape froze in at the distance $15-17 R_E$ on an orbit with $e \simeq 0.2$ \cite{Keane:2014}. 

Our high-obliquity model is currently unique in explaining the origin of large past lunar inclination, which was subsequently reduced by strong obliquity tides at the Cassini state transition \cite{Chen:2016}. A high-obliquity early Earth offers an angular momentum loss mechanism more robust than the evection resonance \cite{Cuk:2012, Wisdom:2015}. Therefore our results support high angular momentum giant impact scenarios for lunar origin\cite{Cuk:2012, Canup: 2012, Lock:2016}. An initially high-obliquity Earth is consistent with the expectation of random spin axis orientations for terrestrial planets after giant impacts, and the dynamics discussed here naturally reduces Earth's obliquity to low to moderate values. This mechanism also provides a new route by which initially highly tilted terrestrial exoplanets can acquire low obliquities and potentially stable climates.

\bibliographystylemp{naturemag}
\bibliographymp{refs1}{}

\begin{addendum}
 \item[Acknowledgments] This work was supported by NASA's Emerging Worlds program, award NNX15AH65G.
 \item[Author Contributions] M.{\'C}. designed the study, wrote the software, analyzed the data and wrote the paper. Co-authors contributed ancillary calculations, discussed the results and wider implications of the work, and helped edit and improve the paper. 
 \item[Author Information] Reprints and permissions statement is available at www.nature.com/reprints. The authors declare that they have no competing financial interests. Correspondence and requests for materials
should be addressed to M.{\'C}.~(mcuk@seti.org).
\end{addendum}

\begin{figure}
\begin{center}
\includegraphics{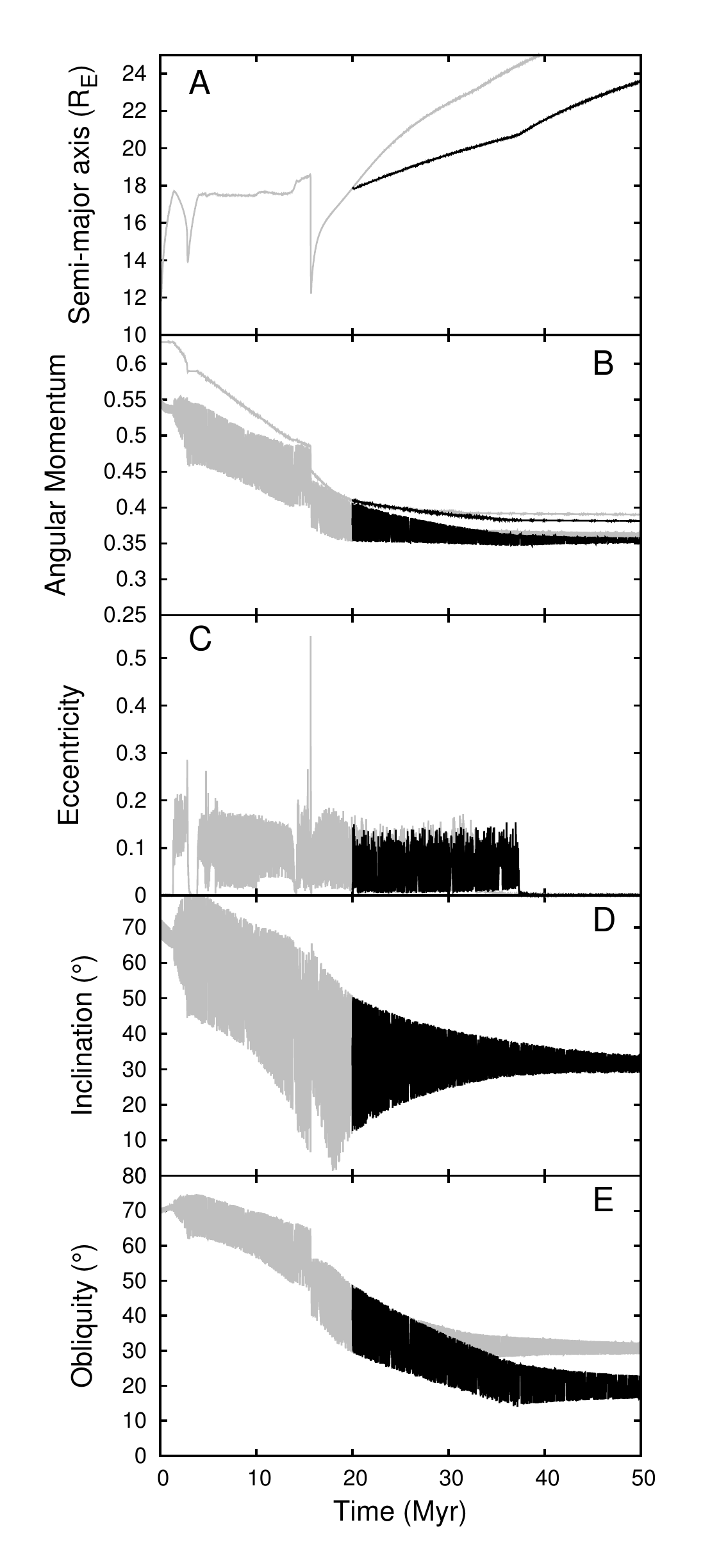}
\end{center}
\end{figure}

\setcounter{figure}{0}

\begin{figure}	
\caption{(Previous page) Numerical simulation of the Moon's early tidal evolution from Earth with initial obliquity of 70$^{\circ}$ and spin period of 2.5~h. The panels show semimajor axis (A), eccentricity (C) and inclination to the ecliptic (D) of the lunar orbit, as well as Earth's obliquity to ecliptic (E) and the angular momentum (AM) of the system (B) in units of $\alpha_E \sqrt{G M_E^3 R_E}$, where $M_E$, $R_E$ and $\alpha_E=0.33$ are the mass, radius and the scaled moment of inertia of Earth, respectively, and $G$ is the gravitational constant. In these units, the present AM is 0.35. The gray lines plot a simulation in which tidal properties of Earth and the Moon were $k_{2,E}/Q_E = k_{2,M}/Q_M = 0.01$ throughout (see SI Section 1 for definitions). The black line shows a simulation branching at $20$~Myr by changing $k_{2,E}/Q_E$ to 0.005. In the AM plot, the thin lines plot a scalar sum of spin and orbital AM, while the lower (thick) band includes only the component of lunar orbital AM vector perpendicular to the ecliptic.}
\end{figure}

\begin{figure}
\begin{center}
\includegraphics[scale=1.]{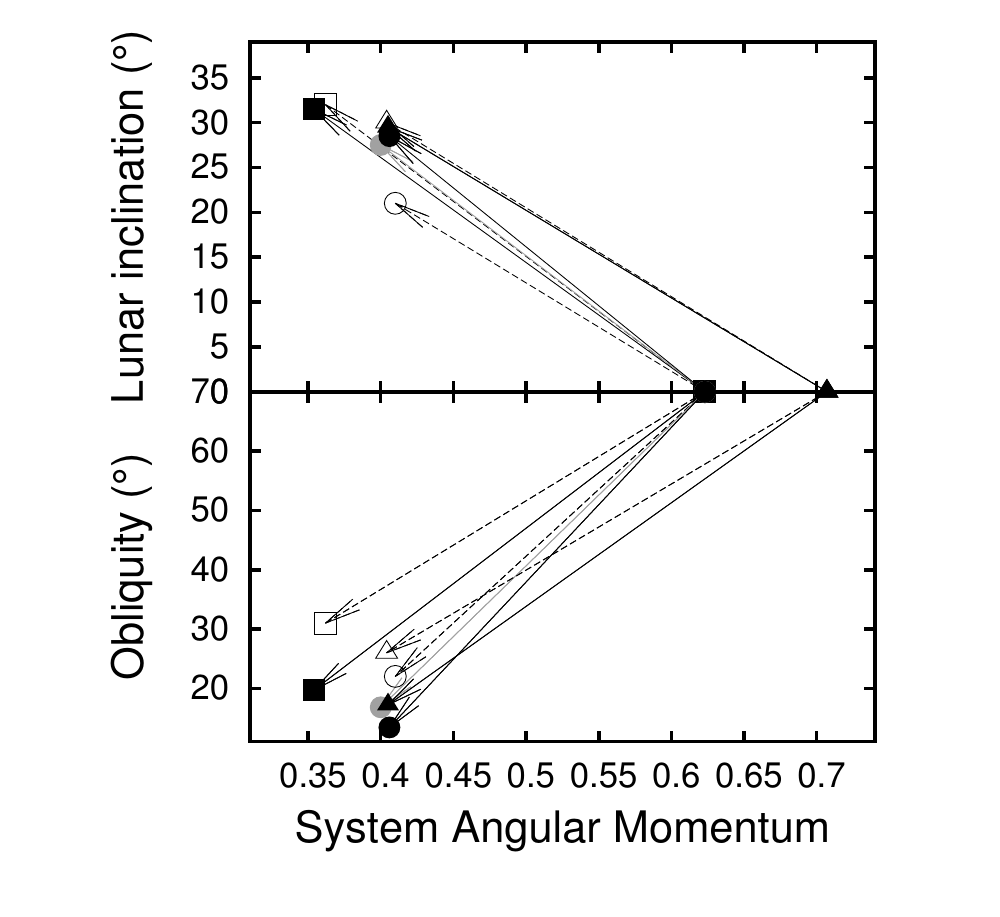}
\end{center}
\caption{Initial and final lunar inclination (top panel) and Earth's obliquity (bottom panel) plotted against corresponding Earth-Moon system angular momenta in different simulations. Angular momentum is plotted in units of $\alpha_E \sqrt{G M_E^3 R_E}$, and all simulations of the Laplace plane transition included here started with Earth's obliquity of $70^{\circ}$. The squares correspond to simulations plotted in Fig. 1: open square plots simulation with $Q_E/k_{2,E}=Q_M/k_{2,M}=100$ while the filled square plots the branch for which $Q_E/k_{2,E}=200$ after 20 Myr. Circles plot simulations with $Q_E/k_{2,E}=200$ throughout (shown in ED Fig. 5), with white, gray and black circles corresponding to cases with $Q_M/k_{2, M}=200$, 100, and 50, respectively. Simulations with higher initial AM (2.25 hr initial spin) (ED Figure 6) are plotted with triangles: the open triangle plots a simulation with $Q_E/k_{2,E}=Q_M/k_{2,M}=100$ while the filled triangle plots the branch for which $Q_E/k_{2,E}=200$ after 30 Myr. All simulations saw a significant loss of AM, excitation of lunar inclination to large values, and a large reduction of Earth's obliquity. We conclude that a smaller $Q_E$ leads to a larger AM loss, while a greater $Q_E$ later in the Laplace plane transition leads to smaller final obliquity for Earth.}
\end{figure}

\begin{figure}
\begin{center}
\includegraphics[scale=1.]{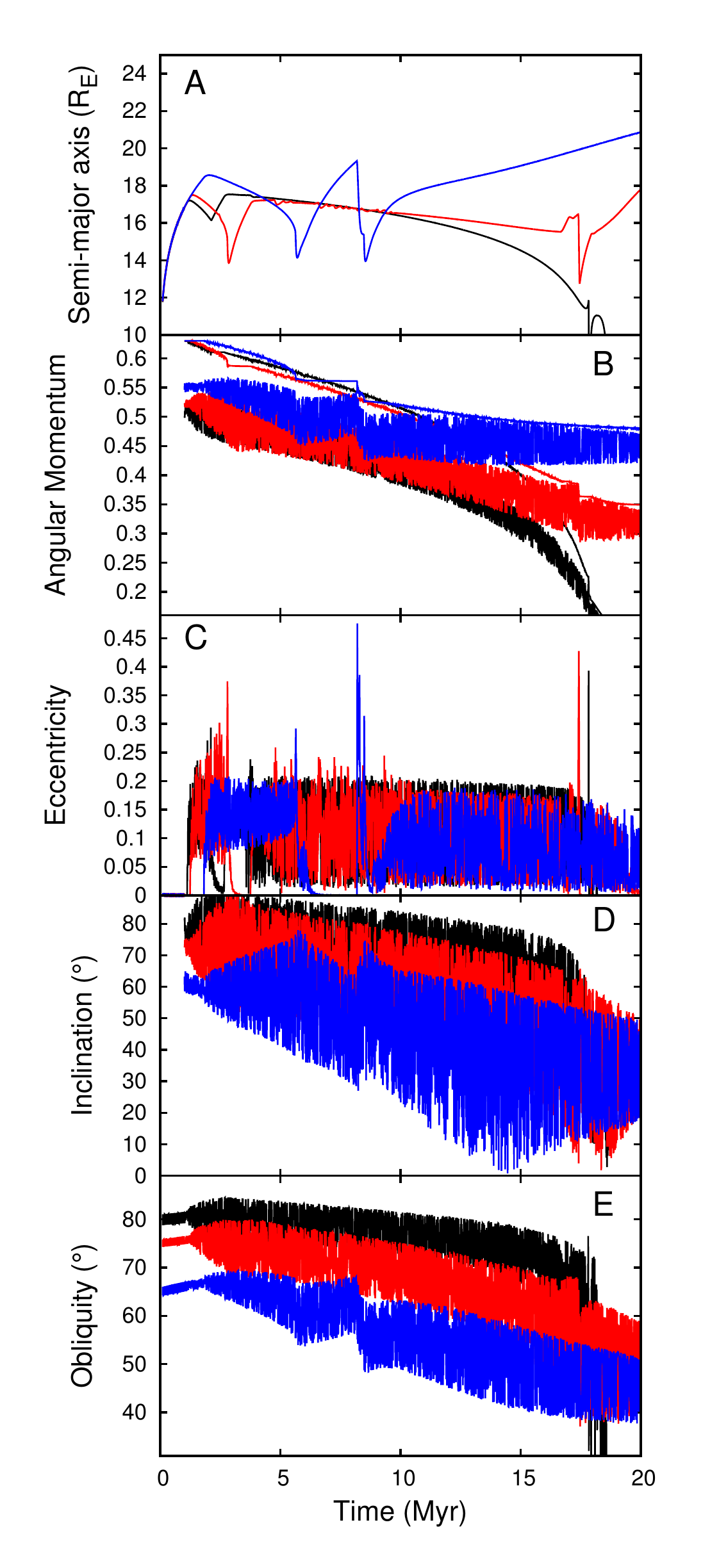}
\end{center}
\end{figure}

\setcounter{figure}{2}

\begin{figure}
\caption{(Previous page) Similar to Fig. 1, but with different initial obliquities for Earth: 80$^{\circ}$ (black line), 75$^{\circ}$ (red line) and 65$^{\circ}$ (blue line). $Q_E/k_{2,E}=Q_M/k_{2,M}=100$ throughout the simulations. Unlike the simulations shown in Figs. 1, these simulations were followed for only 20 Myr. The 80$^{\circ}$ simulation leads to the Moon falling back on Earth. We note a trend that the angular momentum loss (panel B) is larger for higher initial obliquities of Earth.}
\end{figure}

\begin{figure}

\caption{(Next page) Numerical integration of the later phase of lunar tidal evolution, assuming a lunar inclination of 30$^{\circ}$ at 25~$R_E$ and the current shape of the Moon. The panels plot lunar inclination to the ecliptic (A), lunar obliquity with respect to its orbit (B) and lunar spin rate against the Moon's semimajor axis. The red points plot the segments of the tidal evolution that were artificially accelerated, while the blue points plot the intervals integrated with nominal parameters. We used $k_{2,E}=0.3$, $Q_E=35$, $k_{2,M}=0.024$ and $Q_{M, 0}=60$ (for the numerical implementation of lunar tides, see Methods Section 3), with the Love numbers enhanced 100x before the event at 29.7~$R_E$ and by 10x after that event. Black lines plot the Cassini state obliquity (in panel B) and the synchronous rotation rate (in panel C) expected for each instantaneous semimajor axis and inclination. The Moon is in non-synchronous rotation from 29.7~$R_E$ to 35~$R_E$.}
\end{figure}

\setcounter{figure}{3}
\FloatBarrier

\begin{figure}
\begin{center}
\includegraphics[scale=1.]{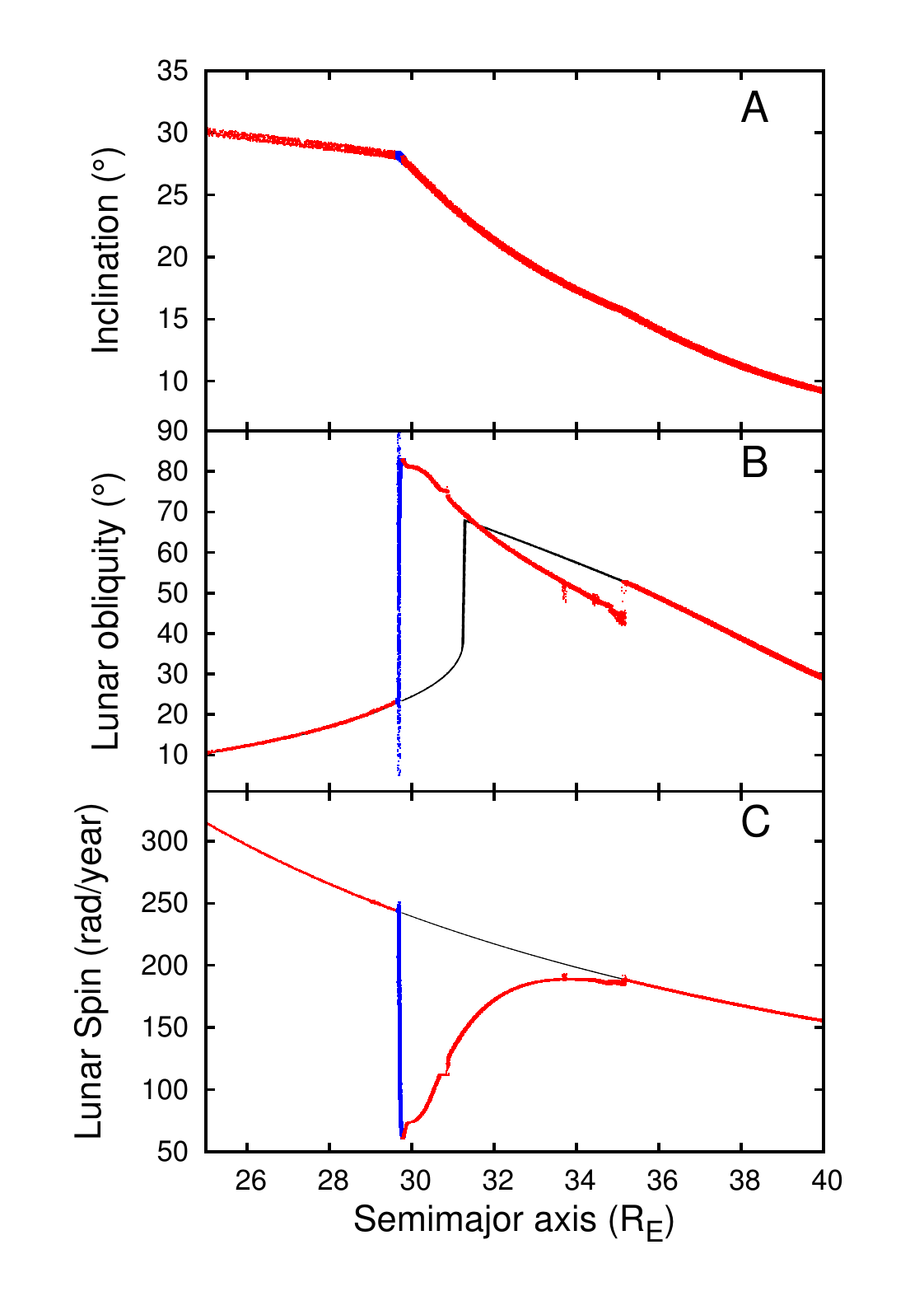}
\end{center}
\end{figure}

\FloatBarrier

\newpage

\renewcommand{\thefigure}{ED\arabic{figure}}
\renewcommand{\thetable}{M\arabic{table}}
\renewcommand{\theequation}{M\arabic{equation}}

\renewcommand{\figurename}{Extended Data, Figure}

\setcounter{equation}{0}
\setcounter{figure}{0}

\section*{Methods}

\paragraph*{1. Damping of lunar inclination by obliquity tides}

In order to quantify the effect of obliquity tides on the lunar orbit, we constructed a semi-analytical method for modeling the evolution of lunar orbit under the influence of Earth and Moon tides. This model assumes that the Moon is in synchronous rotation and the relevant Cassini state, unlike the fully numerical integrator {\sc r-sistem} that we use elsewhere in this work which fully resolves lunar rotation. The goal of this model is to (a) simply demonstrate the importance of lunar obliquity tides during the Cassini state transition, and (b) allow for more efficient integration of the lunar orbit beyond 40 $R_E$, when the Moon was likely in Cassini state 2, and its obliquity, inclination and eccentricity were moderate or small.  

The semi-analytical model is described by a system of equations \cite{Chyba:1989}$^{,}$\citesm{Peale:1978, Canup:1999}:

\begin{equation}
\label{dotae}
\Bigl({da \over dt}\Bigr)_E={3 k_{2,E} M_M \over Q_E M_E} \Bigl({R_E \over a}\Bigr)^5 a n 
\end{equation}

\begin{equation}
\Bigl({de \over dt}\Bigr)_E={19 \over 8 }{e \over a} \Bigl({da \over dt}\Bigr)_E
\label{dotee} 
\end{equation}

\begin{equation}
\Bigl({di \over dt}\Bigr)_E=-{1 \over 4 }{\sin{i} \over a} \Bigl({da \over dt}\Bigr)_E
\label{dotie} 
\end{equation}

\begin{equation}
\Bigl({de \over dt}\Bigr)_M=-e {21 k_{2, M} M_E \over 2 Q_M M_M} \Bigl({R_M \over a}\Bigr)^5 n
\label{dotem}
\end{equation}

\begin{equation}
\Bigl({di \over dt}\Bigr)_M=- {\sin^2{\theta} \over \tan{i}} {3 k_{2,M} M_E \over 2 Q_M M_M} \Bigl({R_M \over a}\Bigr)^5 n
\label{dotim}
\end{equation}

\begin{equation}
\Bigl({da \over dt}\Bigr)_M=2 a e \Bigl({de \over dt}\Bigr)_M +2 a \tan{i} \Bigl({di \over dt}\Bigr)_M
\label{dotam}
\end{equation}
where $a$, $e$, $i$ and $n$ are the semimajor axis, eccentricity, inclination and mean motion of the lunar orbit, respectively; $t$ is time; $M$, $R$, $k_2$ and $Q$ are mass, radius, Love number and tidal quality factor, while the subscripts E and M refer to Earth and Moon, respectively (subscripts after derivatives indicate if the effect is due to Earth or lunar tides). Lunar obliquity to the orbit, $\theta$, is calculated as the solution to the equation \citemp{Ward:1975}$^{,}$\citesm{Peale:1969}:
\begin{equation}
\label{ward}
{2 \over 3} n \Bigl({C-A \over C}\Bigr) \sin{\theta} \cos{\theta} + {3 \over 8} n \Bigl({B-A \over C}\Bigr) \sin{\theta}(1-\cos{\theta}) + \dot{\Omega} \sin(\theta - i) = 0
\end{equation}
where $A$, $B$ and $C$ are the Moon's principal moments of inertia, and $\dot{\Omega}$ is the rate of precession of the lunar node, which we estimate as \citesm{Cuk:2004}:
\begin{equation}
\dot{\Omega}= {3 \over 4} {n^2_E \over n} \cos{i}
\label{node}
\end{equation}
where $n_E$ is Earth's heliocentric mean motion. This expression for $\dot{\Omega}$ assumes precession dominated by solar perturbations and is invalid close to Earth where the influence of its equatorial bulge is important. Our numerical implementation is a simple mapping that solves for lunar obliquity, calculates tidal derivatives and then advances the orbital elements (we used a 1~Myr timestep, appropriate for distances beyond $\sim 25 R_E$). 

\begin{figure}
\includegraphics{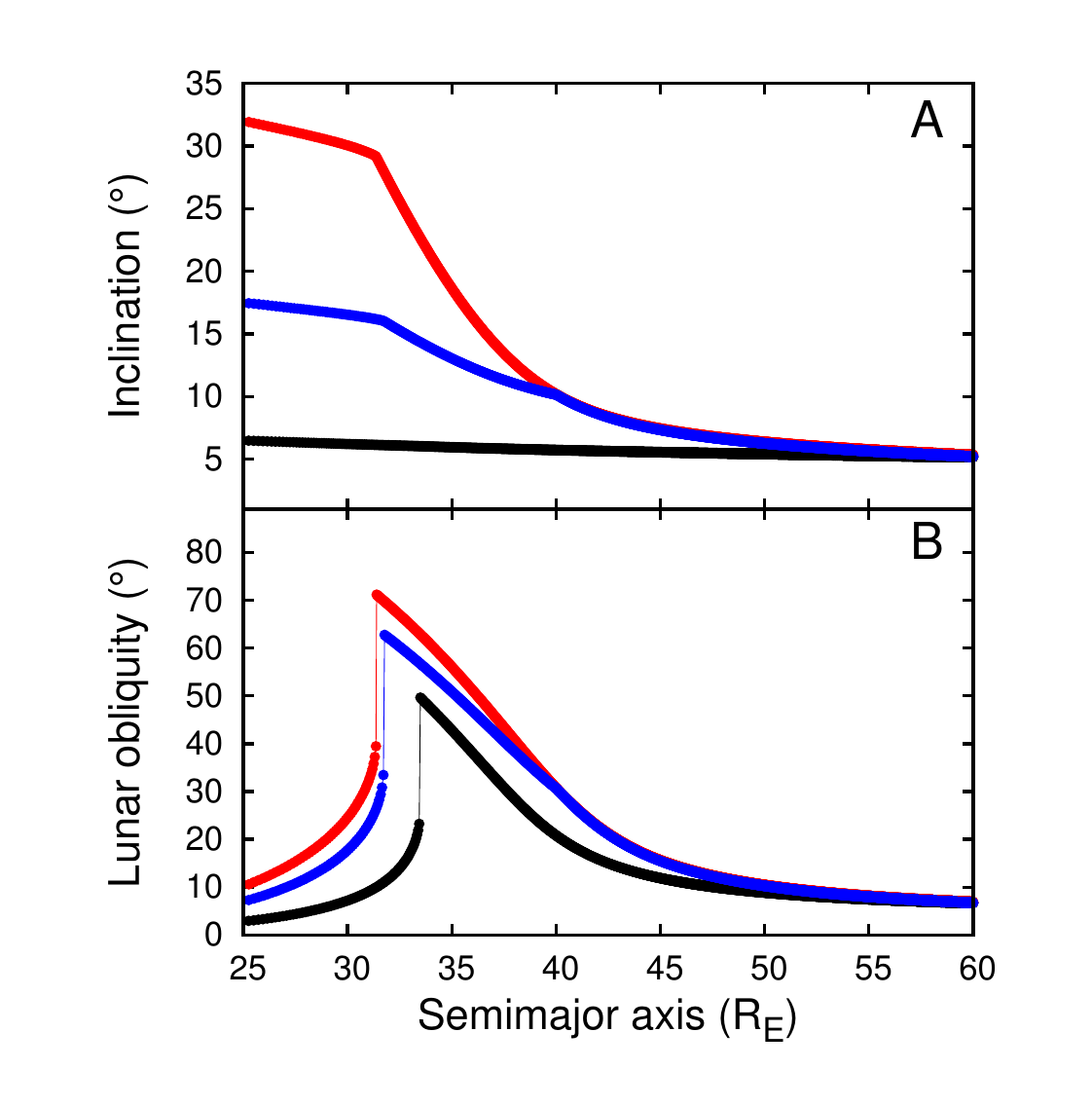}
\caption{\textbf{\textsf{Semi-analytical model of the lunar tidal evolution.}} Evolution of lunar inclination (A) and obliquity (B) as the Moon evolves from 25 to 60~$R_E$ using our semi-analytical model (Methods Section 1). Initial inclinations were chosen so that the final lunar inclination was the current value of about 5$^{\circ}$, while the obliquity was calculated assuming the Moon was in a Cassini state (jumps between 30 and 35~$R_E$ are due to transition between Cassini states 1 and 2)\cite{Ward:1975}. Love numbers were set at their current values ($k_{2, E}=0.3$, $k_{2 M}=0.024$) \cite{Williams:2015}, and the current lunar shape was assumed. The black and red lines plot the solutions for $Q_M=10^4$ and $Q_M=38$ (current value), respectively, while $Q_E$ was in the 33-35 range (it was adjusted so that semimajor axis of $60 R_E$ was reached after 4500 Myr). The blue line plots a history assuming $Q_M=100$ interior to $40 R_E$, and $Q_M=38$ after the Moon passes that distance. The black line closely resembles the results of studies\cite{Goldreich:1966, Touma:1994} that neglected lunar obliquity tides, while the other two curves indicate that the past lunar inclination must have been much larger due to lunar obliquity  tides.\label{hybrid}}
\end{figure}

Figure \ref{hybrid} plots three orbital histories calculated using the above approach that result in the Moon moving from 25 to 60 $R_E$ over 4.5 Gyr, and having a final inclination of about $5^{\circ}$ (the Moon's eccentricity was $\simeq 0.01$ in these calculations). Figure \ref{hybrid} clearly shows that the past studies of lunar tidal evolution that neglected lunar tides \citemp{Goldreich:1966, Touma:1994} greatly underestimate the past inclination of the Moon. Even assuming a lower than current tidal quality factor $Q=100$ for the Moon during the Cassini state transition, the inclination had to be over 17$^{\circ}$ at 25~$R_E$, which would in turn require an inclination of approximately 30$^{\circ}$ when the Moon was close to Earth, much greater than the often-cited $12^{\circ}$ \citemp{Touma:1994}. If we adopt the current tidal parameters of the Moon ($Q=38$), the inclination at 25~$R_E$ was over 30$^{\circ}$, seemingly at odds with formation of the Moon in an equatorial disk.

While our simplified model uses several approximations, we can argue that the values for the past lunar inclination that we calculate are underestimates, for at least three reasons. First, equation \ref{dotim} assumes low obliquities, and at obliquities close to 90$^{\circ}$, the leading coefficient is $15/8$ rather than $3/2$ \citemp{Chyba:1989}, leading to a more rapid damping of inclination. Second, we assumed the current Love number for the Moon ($k_{2, M}=0.024$), which implies that the Moon was as rigid at the Cassini state transition as it is now; as the Moon was only a few hundred Myr old and significantly warmer in the interior at the time of the transition, its Love number would be higher than it is now, again leading to more rapid inclination damping. Third, we assumed constant tidal properties for Earth throughout the calculation, which is in conflict with the fact that the current rate of the Moon's tidal recession is about three times higher than the long-term average \citesm{Williams:2015}. An increase in the tidal dissipation within Earth's oceans over time as suggested by modeling \citesm{Webb:1982, Bills:1999} would mean that the Moon spent more time at the Cassini state transition than we calculated, allowing more damping of inclination. These factors are all independent of the fact that the Moon may not have been in synchronous rotation during the Cassini state transition, as discussed in this work.  

\paragraph*{2. Excitation of lunar inclination by encounters with planetesimals}

Recently, Pahlevan and Morbidelli \citemp{Pahlevan:2015} proposed that the lunar inclination was produced by encounters between the Earth-Moon system and leftover planetesimals. While innovative, this model has several outstanding issues which caused the authors to significantly overestimate the effectiveness of planetesimal encounters in raising the lunar inclination. Pahlevan and Morbidelli \citemp{Pahlevan:2015} do not include lunar obliquity tides explicitly in their model, but assume that very high tidal dissipation rates within Earth moved the Moon to 40~$R_E$ within a few tens of Myr, allowing most of the planetesimal encounters to occur after the Cassini state transition, keeping the newly acquired inclination safe from further damping. This timeline works if $k_{2, E}/Q_E=0.1$ for the early Earth (which we find surprisingly high for a planetary body), but not for Earth's long-term average $k_{2, E}/Q_E=0.01$, which would put the Moon at the Cassini state transition during the epoch of planetesimal encounters. However, given the uncertainties in the tidal properties of early Earth, here we will concentrate instead on the orbital mechanics aspects of the Pahlevan and Morbidelli \citemp{Pahlevan:2015} model.

\begin{figure}
\includegraphics{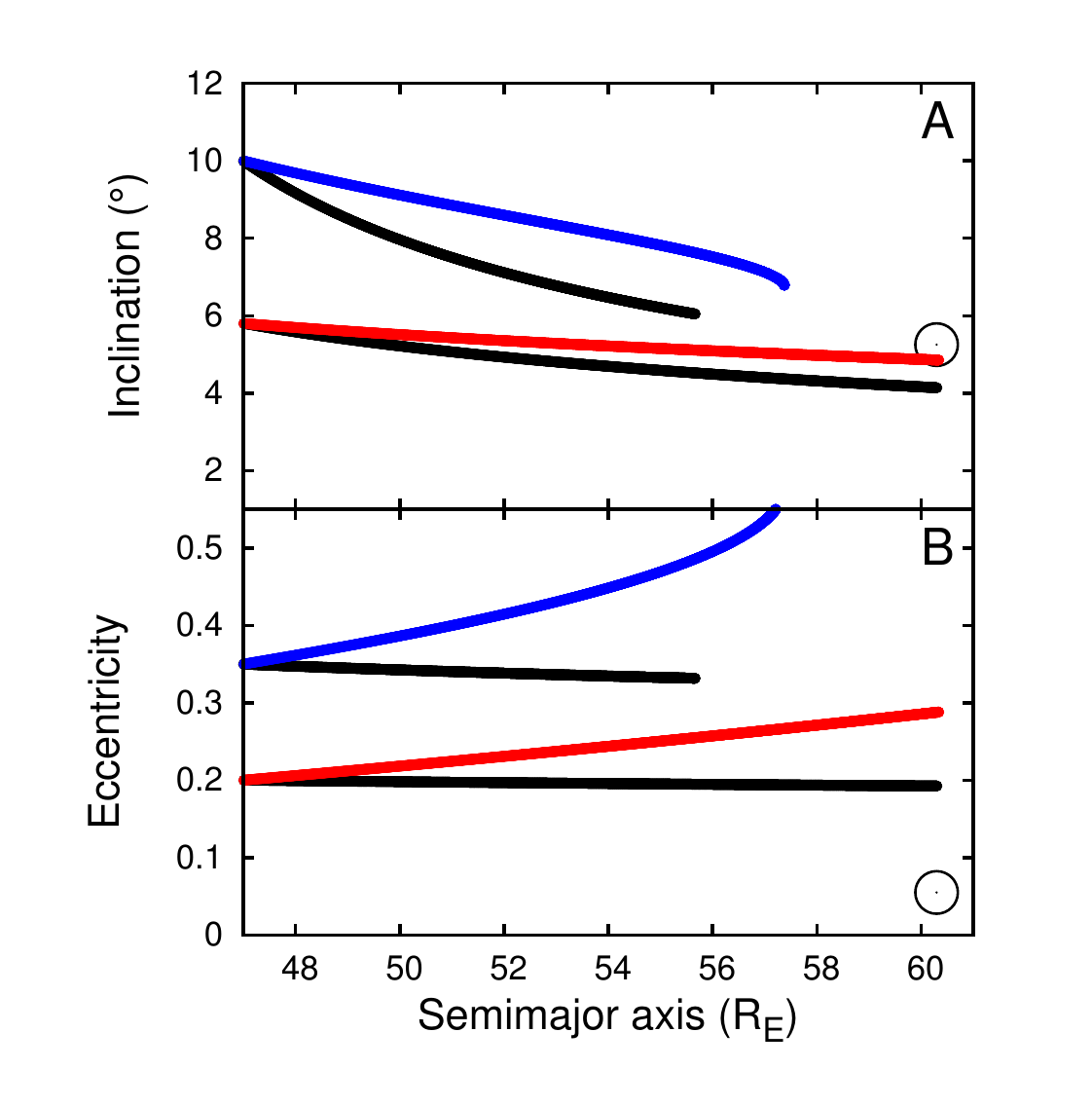}
\caption{\textbf{\textsf{Lunar tidal evolution following planetesimal encounters.}} Evolution of the Moon's inclination (A) and eccentricity (B) following the excitation of the lunar inclination by encounters with planetesimals as proposed by pahlevan and Morbidelli (2015)\cite{Pahlevan:2015}, using our semi-analytical model. The two sets of initial conditions are for the state $a=47 R_E$, $i=5.8^{\circ}$ featured in Pahlevan and Morbidelli (2015)\cite{Pahlevan:2015} Figure 1, and a possible outcome with a more excited inclination $i=10^{\circ}$ (also at $a=47 R_E$). The initial eccentricities were estimated as $e=2 \sin{i}$. The black lines show the evolution assuming $Q_M=38$ and $Q_E=34$, with the current Love numbers. The red line plots the evolution for $Q_M=100$ and the blue line the evolution for $Q_E=20$. The large circular symbol plots the current inclination and eccentricity of the lunar orbit. No combination of tidal parameters can simultaneously match both the current lunar inclination and eccentricity at the same time. A small $Q_M$ combined with high $e$ also keeps the Moon from reaching $60 R_E$ (top two lines). Decreasing $Q_E$ also does not help, as stronger Earth tides further increase the lunar eccentricity (blue line).\label{kaveh}}
\end{figure}

Figure \ref{kaveh} shows the tidal evolution of the Moon using our semi-analytical model (Section S1) starting with the end-state of the simulation featured in Figure 1 of \citemp{Pahlevan:2015} ($a=47 R_E$, $i=5.8^{\circ}$). Eccentricity was not specified in \citemp{Pahlevan:2015}, but it was stated that the encounter simulations typically result in $e=2\sin{i}$, so we have used $e=0.2$. We find that there is, starting from these initial conditions, no combination of tidal parameters for Earth and the Moon that can result in the correct eccentricity and inclination for the Moon at the present epoch. If we assume a larger excitation of lunar inclination, the correspondingly larger lunar eccentricity (given the $e=2 \sin{i}$ condition) counteracts Earth tides and slow down lunar recession. Stronger Earth tides lead to faster outward evolution but also produce a net increase of lunar eccentricity (due to Eq. \ref{dotee}), potentially leading to a reversal of the Moon's orbital evolution through lunar eccentricity tides. A slowdown or reversal in lunar tidal recession would preclude the Moon from reaching its present distance in 4.5 Gyr. Therefore we conclude that for planetesimal encounters to be able to explain the lunar inclination, the encounters must excite inclination without significantly exciting eccentricity (i.e. leave the Moon with lower eccentricity than that reported by \citemp{Pahlevan:2015}). Given the stochastic nature of the process, a small number of encounter outcomes will have a low eccentricity and high inclination. However, while possible, such scenarios are statistically unlikely and therefore not compelling.

Mechanisms other than tides have the potential to alter lunar eccentricity. It has been suggested that the current lunar eccentricity of $0.055$ is the product of a resonance between the Moon and Jupiter that happens when the Moon is at about 53~$R_E$ \citesm{Cuk:2007}. This resonance arises because the rate of precession of the lunar longitude of pericenter is commensurable with the mean motion of Jupiter ("jovian evection"). The rate of precession is (ignoring lunar inclination which is not affected by the resonance) \citesm{Cuk:2004}:
\begin{equation}
\dot{\varpi}= {3 \over 4} {n^2_E \over n} {\sqrt{1-e^2} \over (1-e_E^2)^{3/2}} + {225 \over 32} {n^3_E \over n^2} {{1-e^2} \over (1-e_E^2)^3}+O(n^3_E/n^3)
\label{varpi}
\end{equation}
where $e_E$ is the eccentricity of Earth's orbit. As $e_E$ varies rapidly on 10$^5$-yr timescales due to Milankovi\'c cycles \citesm{Murray:1999}, capture is impossible and the Moon crosses the resonance many times in both directions. These numerous crossings result in a random walk in lunar eccentricity, within a band in $a$ and $e$ associated with the resonance \citesm{Cuk:2007}. Since this band moves to higher $e$ for higher $a$, it is possible to reach arbitrary large eccentricities through this random walk; however it is likely that the Moon will sooner or later reach the outer boundary of the resonant band (defined by the resonance location for $e_E=0$), and exit this chaotic region with finite eccentricity, as shown in \citesm{Cuk:2007}. 

However, it is also possible that through the random walk the lunar orbit will evolve to lower eccentricities. This reduction of eccentricity cannot be arbitrarily effective: it must be smaller than the thickness of the band in $e$ for the semimajor axis at which the Moon enters the resonant band. Keeping $\dot{\varpi}$ and $a$ constant, the thickness of the band in $e$ is determined by the variation in $e_E$:
\begin{equation}
(1-e_{E, max}^2)^{3/2} = {\sqrt{1-e^2_{max}} \over \sqrt{1-e^2_{min}}}
\label{band}
\end{equation} 
where $e_{E, max}$ is the maximum eccentricity of Earth during Milankovi\'c cycles (minimum is zero) and $e_{max}$ and $e_{min}$ are the boundaries of the band at the semimajor axis the Moon encounters the resonant band. Entering the band with $e_{max}$, the Moon cannot acquire eccentricity lower than $e_{min}$. So if we square the Eq. \ref{band} and keep only quadratic terms in eccentricities, we get
$e_{max}^2-e^2_{min} = 3 e_{E, max}^2$ or, assuming that $\Delta e = e_{max} - e_{min} << e_{max}$
\begin{equation}
\Delta e = {3 e_{E, max}^2 \over 2 e_{max}}
\end{equation}
Therefore, for $e_{E, max}=0.06$ \citesm{Murray:1999} a starting eccentricity of $e=0.2$ cannot be reduced by the jovian evection by more than about $0.03$, so the jovian evection does not substantially change the implications of Fig. \ref{kaveh}, namely that the lunar eccentricity will likely remain high if it was substantially excited by planetesimal encounters when the Moon was already beyond 40~$R_E$. 

\paragraph*{3. Numerical methods}

Our dynamics code {\sc r-sistem} ("Symplectic Integrator with Solar Tides in the Earth Moon system"; "R" stands for "Rotation") directly integrates both the orbital and rotational motion of the Moon. The Moon is treated as rigid and tri-axial, and experiences tidal accelerations, the effects of Earth's equatorial bulge and solar perturbations. The orbital part of the integrator is a symplectic mixed-variable integrator based on the principles of Wisdom and Holman \citesm{Wisdom:1991}, with the specific implementation taken from Chambers et al. \citesm{Chambers:2002}. The integrator assumes the Moon is on a Keplerian orbit, with all other forces (including solar perturbations) inserted as periodic "kicks" in Cartesian coordinates. The orbital part of {\sc r-sistem} overlaps substantially with the more general-purpose satellite dynamics code {\sc simpl} \citesm{Cuk:2016}, which has been extensively tested. Like {\sc simpl}, {\sc r-sistem} can include other planets (or other satellites of Earth) as perturbers, but neither were present in any of the integrations included in this paper. With no other planets, Earth's orbit was essentially Keplerian, as {\sc r-sistem} ignores back reaction from tides and lunisolar interactions on Earth's heliocentric orbit.

Integration of the Moon's rotation was based on the Lie-Poisson approach of Touma and Wisdom \citesm{Touma:1994b}, with the Moon treated as a tri-axial rigid body torqued by Earth and the Sun. This enables direct modeling of the Moon's axial precession and Cassini states, and allows for capture into spin-orbit resonances, such as synchronous rotation. One issue we had to deal with is suppression of tumbling (i.e. Chandler wobble) which can be triggered by some spin-orbit interactions. We adopted the approach similar to that of  Vokrouhlick{\'y} et al. \citesm{Vokro:2007}, where a torque perpendicular to the angular momentum (AM) vector acts on the AM vector (in the rotating reference frame) to push the AM vector toward the z-axis (axis of the largest principal moment of inertia). The intensity of this torque is adjusted to match the wobble damping timescales predicted by Sharma et al. \citesm{Sharma:2005}. If the AM vector is closer to the long axis of the body (i.e. that associated with the minimum moment of inertia), a similar torque pushes the AM vector away from the long axis. These torques have reversed signs in opposite hemispheres, so that the wobble damping is the same regardless of the sense or rotation (in the body-fixed reference frame). 

While the rotation of Earth is not resolved in our model, Earth's rotation period is tracked as it is being changed by tides. The oblateness of Earth is adjusted at the end of each timestep depending on the new value of the spin rate. We assumed that $J_2 \sim \omega^2_E$ holds for all spin periods (where $J_2$ is the standard oblateness moment and $\omega_E$ is Earth's spin rate), which may be inaccurate if Earth was very non-spherical due to fast spin. Earth's axis is made to precess due to instantaneous torques by the Moon and the Sun on the equatorial bulge (i.e. we assume that Earth is purely oblate and in principal axis rotation). 

The most important part of the numerical approach are the tides. {\' C}uk and Stewart \citemp{Cuk:2012} approximated lunar tides as a radial force counteracting the radial motion of the Moon. While this implementation could match eccentricity damping (assuming that the Moon is in synchronous rotation), it is not useful in modeling obliquity tides (as a radial force is always in the plane of the orbit). We still include a radial lunar tide, but it only accounts for the damping expected from the actual, physical radial tide (as opposed to libration and obliquity tides). The expression for the radial force used in the integrator is:
\begin{equation}
F_r=-{\dot r} \ {9 k_{2, M} M_E \over Q_M M_M} \sqrt{G M_E \over r^3}  \Bigl({R_M \over r}\Bigr)^5
\label{radial}
\end{equation}
where $r$ and $\dot{r}$ are the Moon's geocentric distance and radial velocity. Using the expression for $\dot{r}$ \citesm{Murray:1999}:
\begin{equation}
\dot{r}= {n a \over \sqrt{1-e^2}} e \sin{f} 
\label{rdot}
\end{equation}
where $f$ is the Moon's true anomaly, and the expression for the eccentricity damping by generalized accelerations \citesm{Burns:1976, Murray:1999}:
\begin{equation}
{de \over dt} = \sqrt{a (1-e^2) \over G M_E } [F_r \sin{f} + F_t (\cos{f}+\cos{E})]
\label{burns}
\end{equation}
where $F_t$ is a tangential acceleration (not present here) and $E$ the eccentric anomaly, we can write
\begin{equation}
{de \over dt} = - e {9 k_{2, M} M_E  \over Q_M M_M} n a^{3/2} R_M^5 r^{-13/2} \sin^2{f}
\end{equation}
To the lowest order in eccentricity, this expression averages over an orbital period to:
\begin{equation}
{de \over dt} = - e {9 k_{2, M} M_E  \over 2 Q_M M_M} \Bigl({R_M \over a }\Bigr)^{5} n
\end{equation}
This is the analytical expression for the eccentricity damping expected from radial tides for a moon in synchronous rotation \citesm{Murray:1999}. While our radial force approximates the predicted damping of eccentricity to lowest order, our approach is fundamentally different from that taken in analytical derivation. While \citesm{Murray:1999} show that energy loss is expected during a monthly cycle of tidal stresses, our integrator ``does not know" about the cycle repeating monthly. All of the forces acting in our code are based only on current positions and velocities (linear or angular) of the Sun, Earth and Moon. Despite these differences in approach, our force form has a clear physical meaning: energy is lost when a somewhat inelastic Moon moves radially in Earth's tidal field, as the mechanical energy going into elastic deformation turns into heat. Since our force is strictly radial, it does not affect the AM of the lunar orbit, in agreement with previous treatments of satellite tides \citesm{Murray:1999}.

We have also applied the principle that only the instantaneous quantities can be used to calculate tidal forces due to librational and obliquity tides. They are not treated differently, but are combined in the same "kick", as they both arise from the Moon's orientation moving relative to the Earth-Moon line. An equal and opposite torque is applied to the Moon's spin, enabling despinning and damping into the synchronous rotation. Our approach is based on analogy with the general case of tides on a non-synchronous body; if we assume that the Moon is rotating much faster than it orbits the Earth, the acceleration experienced by the Moon from tides raised by Earth on the Moon would be \citesm{Murray:1999}:
\begin{equation}
F_t={\rm sign}(\dot{\phi}-n){3 k_{2,M} M_E \over 2 Q_M M_M} \Bigl({R_M \over a}\Bigr)^5 a n^2
\end{equation}
where ${\dot \phi}$ is the lunar spin rate. We may be tempted to apply this equation to the synchronous case, and assume that the lag-angle is independent of synodic frequency (equivalent to having constant $Q$) so that the angular acceleration simply changes sign when the rotation is slower than orbital motion \citesm{Murray:1999}. We can estimate the effect of such tidal acceleration on eccentricity (with obliquity set to zero), using Eq. \ref{burns}:
\begin{equation}
{de \over dt}= (\dot{\phi}-\dot{f}) {3 k_{2,M} M_E \over 2 Q_M M_M} \Bigl({R_M \over a}\Bigr)^5 n \sqrt{1-e^2} \  (\cos{f}+\cos{E})
\end{equation}
For small eccentricities and synchronous rotation, we can assume that $\dot{\phi}-\dot{f}= - 2 e n \cos{M}$, ($M$ being the mean anomaly) so we need to integrate the above expression over $M=[-\pi/2 , \pi/2]$ with a negative sign, and over $M=[\pi/2, 3\pi/2]$ with a positive sign. That would give us an expression for eccentricity damping that does not depend on eccentricity, which is certainly nonphysical. This is because strict constant-$Q$ tides behave like a step function at the exact synchronous motion, without any sensitivity to the amount of deviation from synchronicity.

However, if we assume that the tidal $Q$ does depend on frequency, as $Q^{-1}=Q^{-1}_0 2 (\dot{\phi}-\dot{f})/n$ and keep only the lowest terms in eccentricity, we get:
\begin{equation}
{de \over dt} = - 6 e {k_{2,M} M_E \over Q_M M_M} \Bigl({R_M \over a}\Bigr)^5 n  
\end{equation}  
which is the correct form for the eccentricity damping by the libration tide within a synchronous satellite \citesm{Murray:1999}. Therefore, in order to have a unified treatment of tides in synchronous (or near-synchronous) and non-synchronous cases, we need to assume that the lag angle is proportional to the angular velocity of libration $\dot{\phi}-\dot{f}$ when the deviation of the rotation rate from orbital motion is smaller than the orbital motion, and assume the lag angle is constant when the rotation is much faster than orbital motion. An alternative of constant time-lag tides (for all frequencies) is possible but leads to very large lag angles and very fast tidal accelerations when early Earth had a fast spin \citesm{Meyer:2010, Meyer:2011}; our even higher AM initial conditions would make this problem worse. 

In order to treat tides on the Moon and Earth uniformly, including the obliquity-related terms, we implemented the following relationship between the tidal $Q$ (inverse of the lag angle) and frequency:
\begin{equation}
Q=Q_0 \sqrt{1+{1 \over \delta^2}-{1 \over \delta}}
\end{equation}
with the parameter $\delta$ defined as:
\begin{equation}
\delta=|\mathbf{w} - \mathbf{v}_t|/|\mathbf{v}_t|
\end{equation}
where ${\mathbf v}_t$ is the tangential component of the perturber's apparent velocity relative to the perturbee's (i.e. the deformed body's) center of mass, and the vector ${\bf w}$ is defined as ${\mathbf w} = {\mathbf s} \times {\mathbf r}$, with ${\mathbf s}$ being the angular velocity vector of the perturbee, and ${\mathbf r}$ the radius-vector of the perturber relative to the perturbee. This gives us $Q=Q_0/\delta$ for small librations, and $Q=Q_0$ for a fast-rotating body, and a reasonable transition when $\delta \simeq 1$. The way this lag angle is implemented in the integrator is to place a prolate quadrupole moment on the perturbee, with the axis of symmetry defined by the vector \boldmath${\hat r}+{\hat \delta}$\unboldmath$(2Q)^{-1}$ (hats denote unit vectors). This quadrupole moment (described by Eq. 4.145 of \citesm{Murray:1999}) then torques (and is torqued by) the perturber, producing tidal accelerations. On the Moon we have only the tidal bulge raised by Earth; on Earth we have both lunar and solar tidal bulges, each of which torques both the Sun and the Moon, as lunisolar tidal cross-terms have been found to be important \citemp{Touma:1994}$^{,}$\citesm{Laskar:2004}.

Since we treated all orientations of relative motions between the orbital motion and the perturbee's velocity equally, this formulation does not discriminate between (eccentric) libration tides and obliquity tides. Our tests show that the integrator correctly reproduces the expected relationship \citemp{Chyba:1989}
\begin{equation}
{di \over dt} = {1 \over 7} {\sin^2{\theta} \over e \tan{i}} {de \over dt}
\end{equation}

Our ``leveling off" of the tidal phase lag for libration rates equal to about half of the orbital motion has little bearing on eccentricity evolution, as lunar eccentricities in our scenario are rarely above 0.2. However, we do encounter large obliquities around the Cassini state transition, which means that we may be underestimating inclination damping at high obliquities when compared to analytical calculations \citesm{Peale:1978}$^{,}$\citemp{Chyba:1989}. We think that our approach is justified, as stronger dissipation at large obliquities would require tidal lag angles much larger than that corresponding to the tidal $Q$ in the fast-rotation case, which we treat as the upper limit on the lag angle. Further work may be needed to reconcile different definitions of tidal $Q$s in different treatments of high obliquity tides. 

\paragraph*{4. Laplace plane transition}

In Fig. 1, where a fast-spinning Earth's obliquity was set to $70^{\circ}$, we observe the instability associated with the Laplace plane transition when the Moon is at $17-18 R_E$. During this transition the Moon acquires large inclination, which is similar to the findings of Atobe and Ida \citemp{Atobe:2007}. We find that the behavior of lunar eccentricity is rather complex, exhibiting spikes and crashes, separated by relatively stable periods of moderately excited eccentricity. In order to understand these quasi-stable states, we looked in detail at the results of the simulation at 12.8 Myr (Fig. \ref{frame}). It is clear that the eccentricity is excited by Kozai-type perturbations \citesm{Lidov:1962}$^{,}$\citemp{Kozai:1962}, which are secular and involve interaction between an outer perturber (the Sun in this case) and an inclined perturbee (the Moon). Unlike in standard Kozai dynamics of planetary satellites \citesm{Innanen:1997, Carruba:2002}, Earth's obliquity plays an important role, and the inclination oscillates due to mutual precession of the lunar orbit and Earth's spin axis, so that the inclination and eccentricity are not simply anti-correlated. Fig. \ref{frame} shows that this dynamical state is periodic in its secular behavior, with exactly three inclination cycles for every eccentricity cycle. We hypothesize that each of the stable intervals in eccentricity (Figs. 1, 3, \ref{slow} and \ref{spin}) represent a periodic secular state where certain precession periods are locked in resonance. 

Fig. \ref{frame3} shows a different ``slice" of the simulation from Fig. 1 for the $Q_E/k_{2,E}=200$ branch of the simulation (black line) at 34.6~Myr. At this time, inclination is about $30^{\circ}$ and the Laplace plane is dominated by the Sun. The eccentricity is still excited, but not by Kozai-type perturbations. The source of eccentricity excitation are near-resonant perturbations stemming from the slow-varying argument $\Psi=3 \Omega + 2 \omega - 3 \gamma$, where $\Omega$ and $\gamma$ are the longitudes of the lunar ascending node and Earth's vernal equinox, respectively, and $\omega$ is the Moon's argument of perigee. Since this interaction term combines solar perturbations and the precession of Earth's spin axis, it is capable of changing the obliquity of Earth. Since this interaction does not require capture into resonance, it is probably less sensitive to random perturbations such as planetesimal encounters \citemp{Pahlevan:2015} than a narrow resonance would be. We hope to study dynamics of this and similar near-resonant terms in future work. 

Figure \ref{slow} shows the evolution of the Earth-Moon system with the same initial conditions as in Fig. 1, but with $Q_E/k_{2,E}=200$ from the beginning of the simulation. The black line plots the case with $Q_M/k_{2,M}=200$, and the grey line represents $Q_M/k_{2,M}=50$. We also ran a simulation with $Q_M/k_{2,M}=100$, and it was in all ways intermediate between these two. Overall, the simulations shown in Fig. \ref{slow} have low final obliquities, but have an excess of AM compared with the real Earth-Moon system. We conclude that higher $Q_E$ during the early part of the Laplace plane transition leads to less AM loss than in the case shown in Fig. 1, but that it leads to larger reduction in obliquity during the later part of the simulation. Lunar tidal properties appear to be less important, with a more dissipative moon leading to slightly more AM loss and lower final obliquities for Earth.

In Fig. \ref{spin} we started Earth with a spin period of only 2 h (as opposed to 2.5 h in Fig. 1). Just like in Fig. 1, we changed the tidal properties of Earth halfway through the simulation, leading to two different outcomes (see Fig. \ref{spin} caption). While the qualitative evolution of the system is similar to that shown in Fig. 1, there are some qualitative differences. The final obliquity is lower than in Fig. 1 (in excellent agreement with \citemp{Rubincam:2016}), while the final angular momentum is about 15\% too large.  Given that the trend in outcome we see with increasing initial AM (lower final obliquity and higher final AM) is in the opposite direction from the trend we see with increasing initial obliquity (higher final obliquity and lower final AM), it is likely that a more complete exploration of initial conditions will find higher-AM cases where both the final AM and obliquity are satisfactory.

The apparent threshold obliquity for the instability that we find (between $60^{\circ}$ and $65^{\circ}$) is below the critical obliquity of about $69^{\circ}$ found by \citemp{Tremaine:2009}; we understand that this is because of non-zero initial inclination of the Moon when encountering the Laplace plane transition. In all of our simulations, the Moon acquires a few-degrees of inclination when crossing the inclination resonance just interior to the evection \citemp{Cuk:2012}; this early resonance is weak at low obliquities but strong when Earth's spin axis is tilted. As long as we assume a hydrostatic shape for the Moon, the Moon's obliquity is low and obliquity tides are weak, so this inclination largely survives until the Laplace plane transition.

During the Laplace plane transition a large amount of energy is dissipated in the Moon. For example, during the evolution shown in Fig. 1, tidal heating in the Moon reaches $10^{14}-10^{15}$~W for several tens of Myr. Due to the prolonged nature of both the transition and subsequent cooling of the lunar mantle, the lunar crust would contain the signal of major thermal events that occurred tens of Myr after lunar accretion; thus, the Laplace plane transition should be considered when interpreting the geochronology of lunar samples.

\paragraph*{5. Cassini state transition}

The rotation states of most large planetary satellites are highly evolved through tidal forces raised by their parent planet. The spin periods of satellites are typically synchronized with the orbital period, and this is true for Earth's Moon. Apart from spin period synchronization, tidal forces also damp any initial, so-called ``free", obliquities of the moons. The most energetically stable final state for satellite obliquities is one where the spin axis maintains a constant angle to the moon's orbital plane. Since both the satellite's spin axis and the orbital plane are precessing (spin axis precesses around the orbital plane, which precesses around the Laplace plane), the most stable final obliquity is one matched to inclination and the precession rates in a way that the spin axis, orbit normal and Laplace plane normal all stay in the same plane during their precession. This arrangement is called a Cassini state, and Cassini states can be calculated using Eq. \ref{ward} \citesm{Peale:1969}. The two solutions relevant here are Cassini state 1, which the Moon will occupy when its spin precession rate is faster than the orbital precession, and Cassini state 2, which is the only possible solution when the spin precession is slower than orbital precession (as is the case at the present day). The Moon is thought to have crossed the Cassini state transition at about $33 R_E$ when the Cassini state 1 ceased to exist and the Moon had to shift to the Cassini state 2 \citemp{Ward:1975}. 

The largest discrepancy between our {\sc r-sistem} integration of the Cassini state transition (Fig. 4) and the semi-analytical solutions (Fig. \ref{hybrid}) is the fact that the Moon is in non-synchronous rotation from $29.7 R_E$ to about $35 R_E$ in the numerical simulation. In order to explore this phenomenon in more detail, we ran 516 short simulations with initial conditions on a grid in $a$ and $i$ covering the Cassini state transition (Fig. \ref{grid}). We found that the Moon settles into a stable sub-synchronous rotation for a wide range of initial conditions with $a=27-37 R_E$. In all of the grid simulations, the Moon was initially non-synchronous, so our results show when the synchronous rotation will not be re-established after being broken, rather than that the synchronous rotation is unstable, in the regions covered with orange Xs in Fig. \ref{grid}. When it comes to intrinsic stability of continued synchronous rotation, we find major differences between Cassini states 1 and 2. Cassini state 1 is by itself stable all the way until its disappearance at $a=31-34 R_E$ (the exact distance depends on inclination), unless disturbed by an outside influence (e.g., a wobble resonance or an impact). On the other hand, we find that Cassini state 2 is intrinsically unstable if its equilibrium obliquity is above $58.15$ (Fig. \ref{obl}), a result well established in the literature \citesm{Beletskii:1972, Gladman:1996}, but not previously relevant for the Moon in the low-inclination case. We also find that once Cassini state 2 becomes stable, the Moon may still occupy a sub-synchronous rotation state just short of synchronous until it is disturbed by a resonance or an impact.

We also find a number of resonances that excite the Moon's rotation and can lead to long term chaos (red crosses in Fig. \ref{grid}). Some of these are likely associated with secondary resonances found by \citesm{Wisdom:2006}. Here we will concentrate on three features seen in Fig. 4: the wobble resonance at $29.7^{\circ}$, the 1:2 spin-orbit resonance at $31^{\circ}$ and several resonances at $34-35^{\circ}$. The wobble resonance at $29.7^{\circ}$ breaks the synchronous rotation by inducing large non-principal axis rotation within the Moon. Fig. \ref{wobble} shows how the wobble amplitude grows as the Moon approaches the resonance (at a tidal evolution rate accelerated 100 times), and Fig. \ref{res} shows the resonance crossing itself, integrated at the nominal tidal evolution rate. This resonance is caused by the commensurability between Earth's heliocentric motion and the lunar libration in longitude, the frequency of which is given by \citesm{Rambaux:2011}:
\begin{equation}
\lambda = \sqrt{3 (B-A) \over C} n
\end{equation}
Currently, the period of the libration in longitude is 1056.1 days \citesm{Rambaux:2011}. As long as the Moon's shape is constant, this period is proportional to the orbital period, so the semimajor axis at which this period is one year is:
\begin{equation}
a_{r}= \Bigl({365.25 {\rm d} \over 1056.1 {\rm d}}\Bigr)^{2/3}  60.3 {R_E} = 29.7 {R_E}
\end{equation}
which is exactly where it occurs in our numerical simulation. After the wobble disrupts synchronous rotation, the Moon settles into a sub-synchronous high-obliquity state, consistent with Figs. \ref{grid} and \ref{obl}. 

At about $31 R_E$, the Moon is briefly captured into a 1:2 spin-orbit resonance. This resonance can be stable at high eccentricities \citesm{Wieczorek:2012}, but we never observed long periods of capture in our high-obliquity, low-eccentricity cases. At $34-35 R_E$ the Moon encounters several resonances which disrupt sub-synchronous rotation and the Moon settles into (by now stable) Cassini state 2. We think that these resonances are related to the 1:3 secondary secular resonance which was found to intersect with the Cassini state 2 at about this geocentric distance \citesm{Wisdom:2006}. Clearly more work is needed to identify these resonant features and explore the full diversity of the Moon's past spin-orbit dynamics. Here we assumed that the Moon already had its present shape at the time of Cassini state transition, which is a reasonable assumption \citemp{Keane:2014} but other shape histories are possible. Also, beyond $30 R_E$ in Fig. 4, lunar tidal evolution is accelerated 10 times over nominal, and the Moon would probably be affected by some of the later resonances more strongly if the evolution was integrated at the nominal rate. Additional factors that we ignored are impacts that can disrupt the Moon's rotational state \citesm{Wieczorek:2012} and the core-mantle interaction which may have been important in generating the ancient lunar magnetic field \citesm{Meyer:2011a}.

Once the Cassini state is reestablished in Fig. 4, we can compare our numerical results with analytical estimates (Eqs. \ref{dotae}--\ref{dotam}). Our numerical code {\sc r-sistem} damps lunar inclination from $16^{\circ}$ at $35 R_E$ to $9.2^{\circ}$ at $40 R_E$. This is slightly below our nominal target of $10^{\circ}$ at $40 R_E$ (based on Fig. \ref{hybrid}), obtained by assuming long-term average Earth and the present Moon beyond $40 R_E$ (Fig. \ref{hybrid}). Also, {\sc r-sistem} damps inclination slightly faster for $Q_{M,0}=60$ than the analytical model (switching to the  analytical approach at $35 R_E$ would give us $10.9^{\circ}$ at $40 R_E$). This is inevitable consequence of the differences between the two approaches, but we are encouraged by the overall convergence of the results. In the absence of any strong constraints on the timeline of the Moon's tidal evolution, we think that the history shown in Figs. 1 and 4 represents the best currently available explanation for the otherwise puzzling lunar orbital inclination.

\clearpage

\bibliographystylesm{naturemag}
\bibliographysm{refs2}{}

\newpage
\pagestyle{empty}

\begin{figure}
\includegraphics{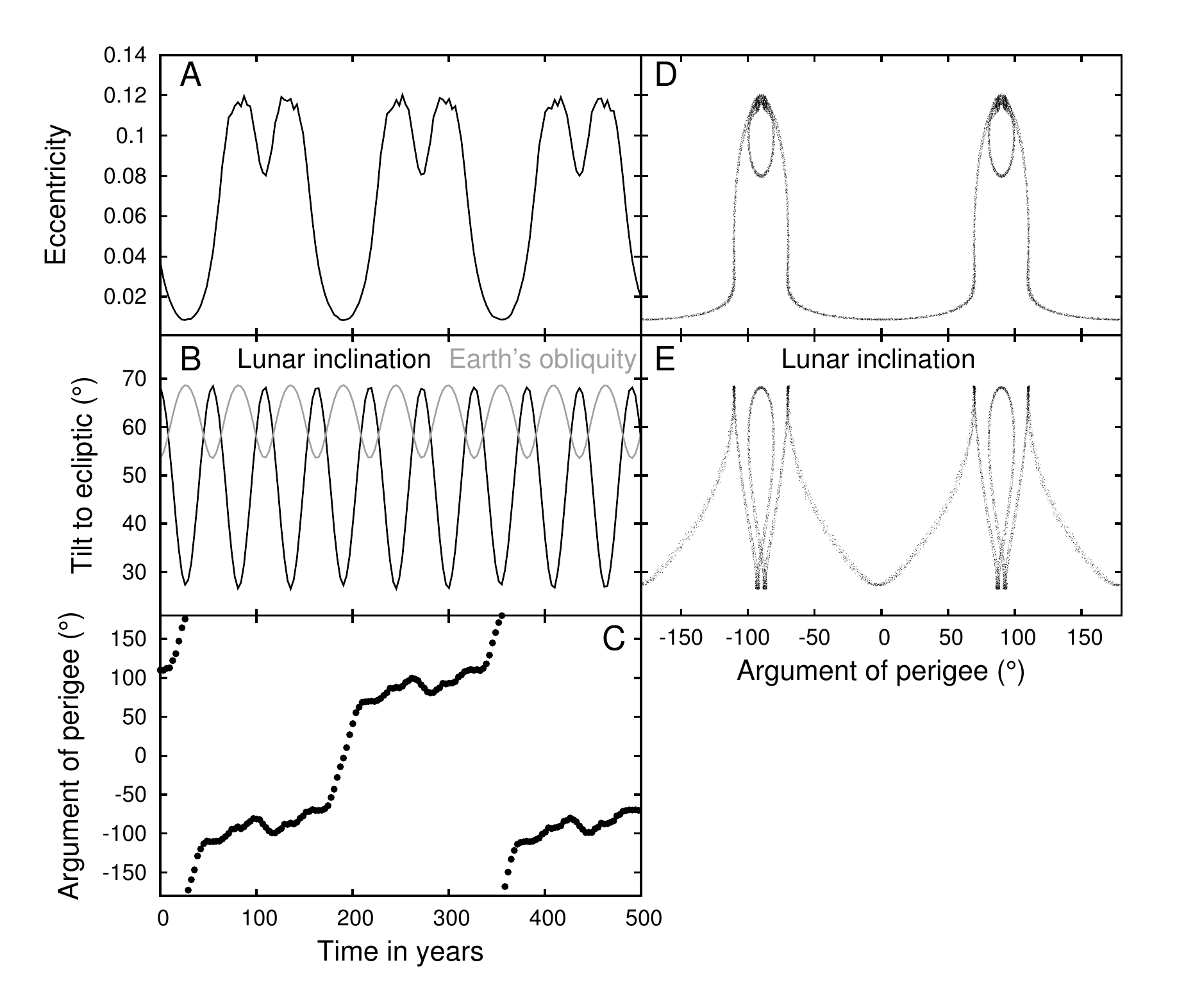}
\end{figure}

\setcounter{figure}{2}

\begin{figure}
\caption{(Previous page) \textbf{\textsf{A snapshot of the simulation shown in Fig. 1 taken at 12.8 Myr.}} The left-hand panels show eccentricity (panel A, top left), lunar inclination and Earth's obliquity with respect to ecliptic (B, middle left), and the argument of perigee of lunar orbit, with the ecliptic as fundamental plane (C, bottom left) versus time over a 500-year period, while the right-hand panels plot lunar eccentricity (D, top) and inclination (E, bottom) against the Moon's argument of perigee (over the whole period of 30,000~yr). The eccentricity is clearly correlated with the argument of perigee, as expected for Kozai-type perturbations \cite{Lidov:1962}$^{,}$\cite{Kozai:1962}. Rapid mutual precession of Earth's spin axis and the plane of the Moon's orbit, which are significantly inclined to one another (and to the ecliptic plane), clearly affects both the eccentricity and the perigee precession. This secular behavior is periodic, with exactly three inclination cycles per one eccentricity cycle, which corresponds to half of the period of precession of the argument of perigee\label{frame}}
\end{figure}

\begin{figure}
\includegraphics{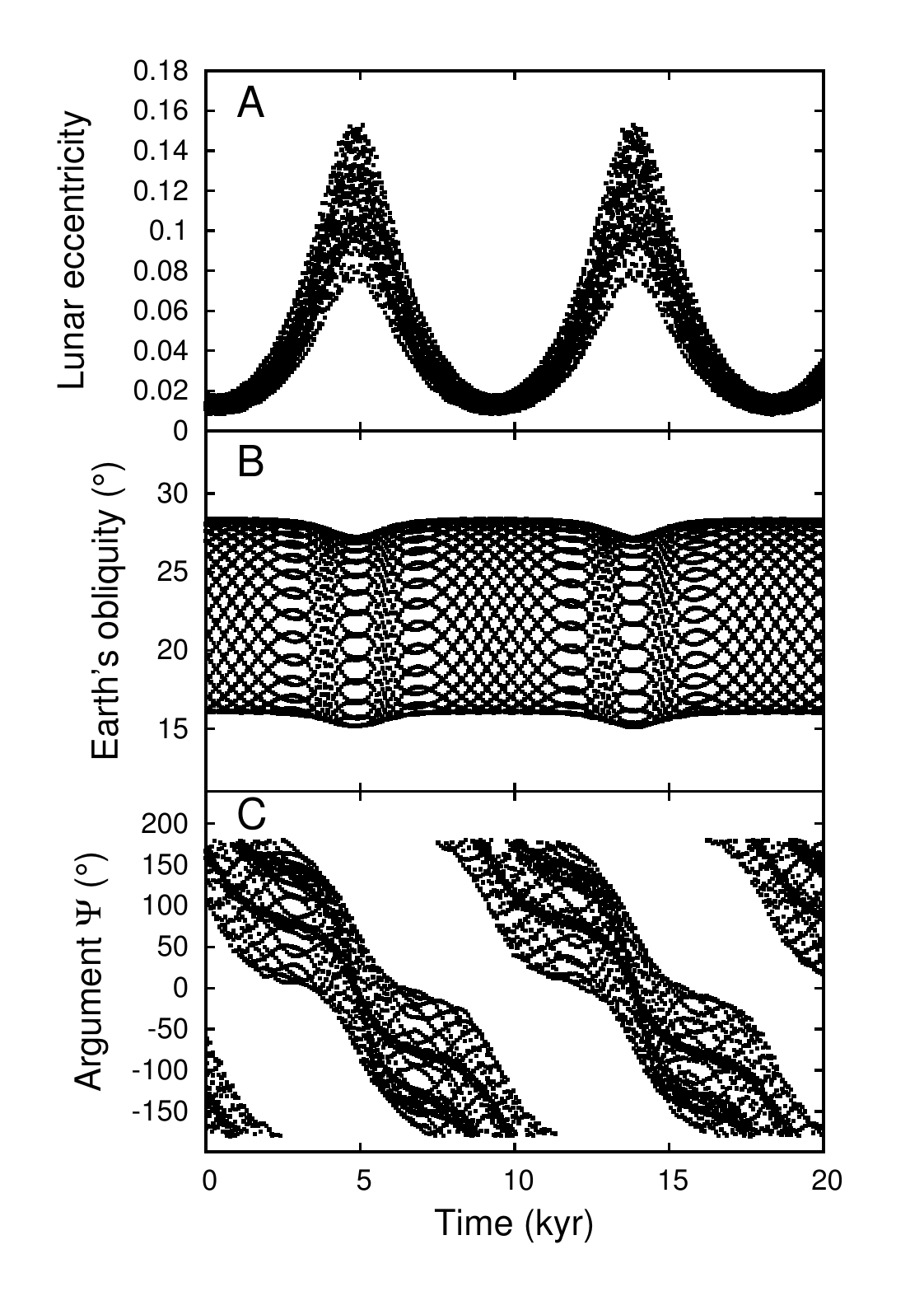}
\caption{\textbf{\textsf{A snapshot of $Q_E/k_{2,E}=200$ simulation shown in Fig. 1 (black line) at 34.6 Myr}}. At this time, eccentricity excitation (top panel) is not due to Kozai perturbations, but due to the slow-varying near-resonant argument $\Psi=3 \Omega + 2 \omega - 3 \gamma$, where $\Omega$ and $\gamma$ are longitudes of the lunar ascending node and Earth's vernal equinox, respectively, and $\omega$ is the Moon's argument of perigee. This near-resonant interaction is responsible for the substantial reduction of Earth's obliquity seen in Fig. 1.\label{frame3}}
\end{figure}
\thispagestyle{empty}

\begin{figure}
\includegraphics{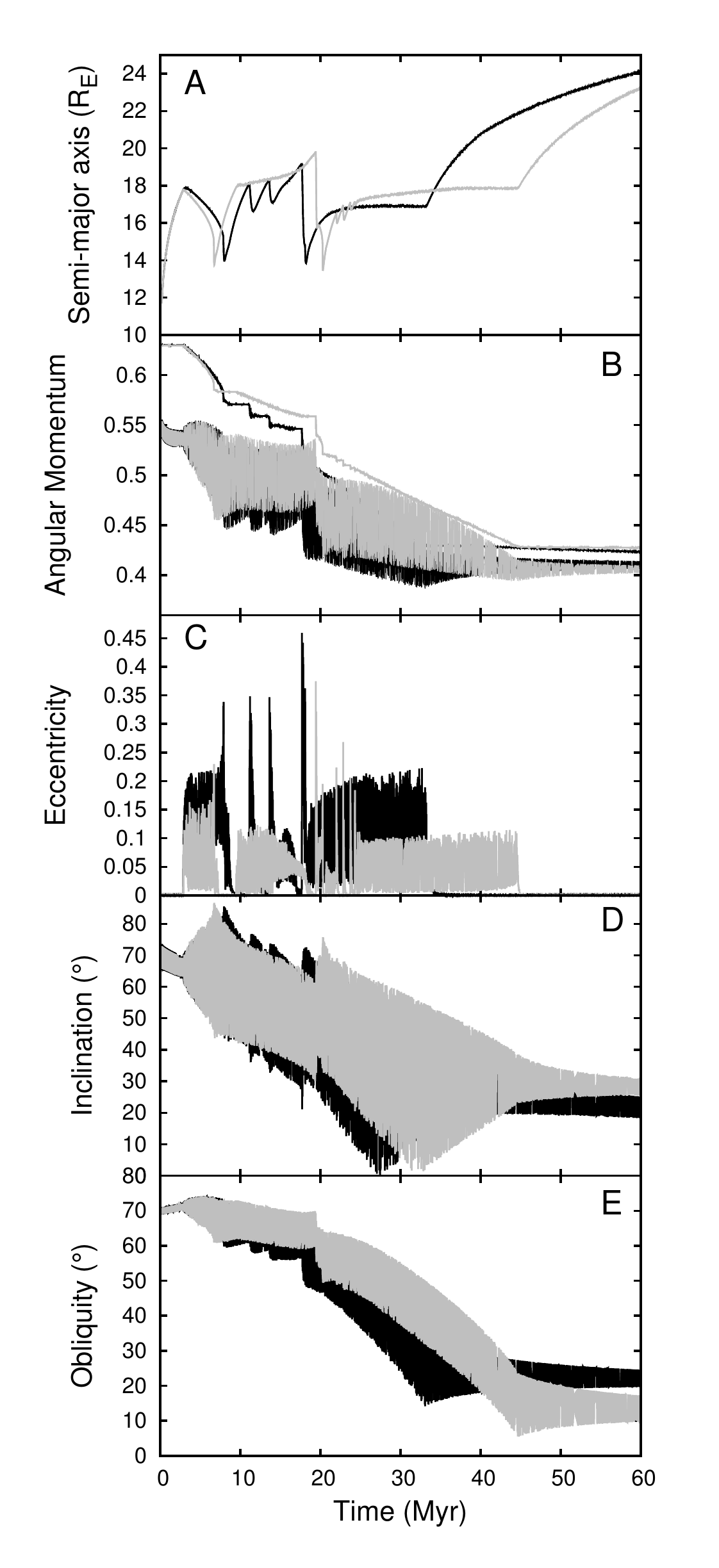}
\end{figure}

\setcounter{figure}{4}

\begin{figure}
\caption{(Previous page) \textbf{\textsf{Early tidal evolution of the Moon with $Q_E/k_{2,E}=200$ throughout the simulations.}} Black lines are for the case with $Q_M/k_{2,M}=200$, while gray lines plot the simulation with $Q_E/k_{2,E}=50$. The most notable aspects of these simulations are low final obliquities of Earth (panel E) and a final angular momentum of the Earth-Moon system (panel B) in excess of the current value of 0.35 $\alpha_E \sqrt{G M^3_E R_E}$.\label{slow}}
\end{figure}

\begin{figure}
\caption{(Next page) \textbf{\textsf{Early tidal evolution of the Moon with Earth initially having a 2 h spin period.}} This is equivalent to the system having twice the current AM. The gray lines plot a simulation in which the tidal properties of Earth and the Moon were $Q_E/k_{2,E} = Q_M/k_{2,M} = 100$ throughout. The black line shows a simulation branching at $30$~Myr by changing $Q_E/k_{2,E}$ to 200. While the final obliquity of Earth (panel E) is correct, the final angular momentum of the Earth-Moon system (panel B) somewhat exceeds the current value of 0.35 $\alpha_E \sqrt{G M^3_E R_E}$.\label{spin}}
\end{figure}

\setcounter{figure}{5}

\begin{figure}
\includegraphics{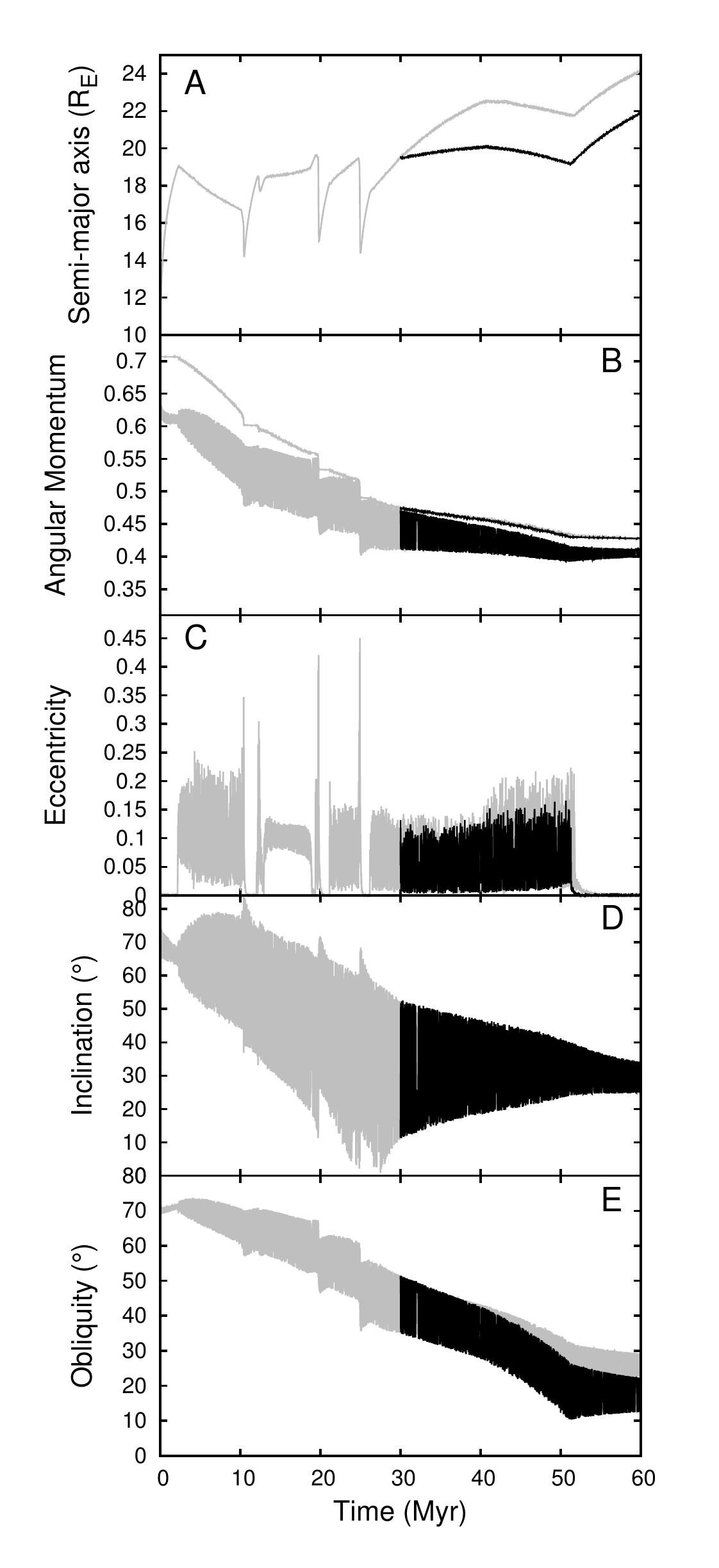}
\end{figure}

\begin{figure}
\includegraphics{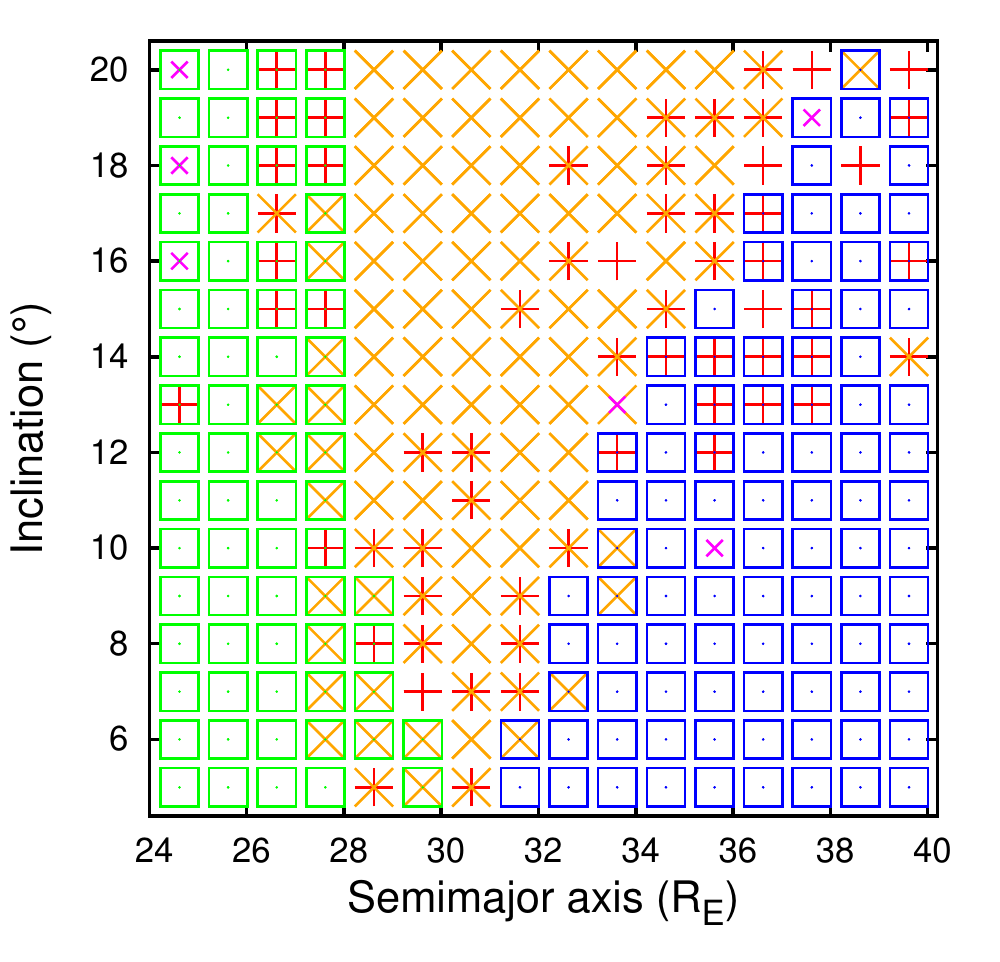}
\caption{\textbf{\textsf{Map of lunar rotational dynamics close to the Cassini state transition.}} Outcomes of 512 simulations probing the end states of initially very fast and very slow lunar rotations for 16 different lunar semimajor axes and 16 different lunar inclinations. Simulations were run for 1 Myr, except for the rightmost three columns, which were followed for 3 Myr. Each $a-i$ field is described by two symbols, one each for initial rotations of 127 rad/yr and 381 rad/yr. Green and blue boxes indicate synchronous rotation in Cassini states 1 and 2, respectively. Xs indicate non-synchronous rotation with stable obliquity, with large orange symbols indicating sub-synchronous, and small magenta Xs plotting super-synchronous states. Red crosses signify variations in obliquity above $1^{\circ}$ during the last 50 kyr of the simulation (indicating excited or chaotic spin axis precession).\label{grid}}
\end{figure}

\begin{figure}
\includegraphics{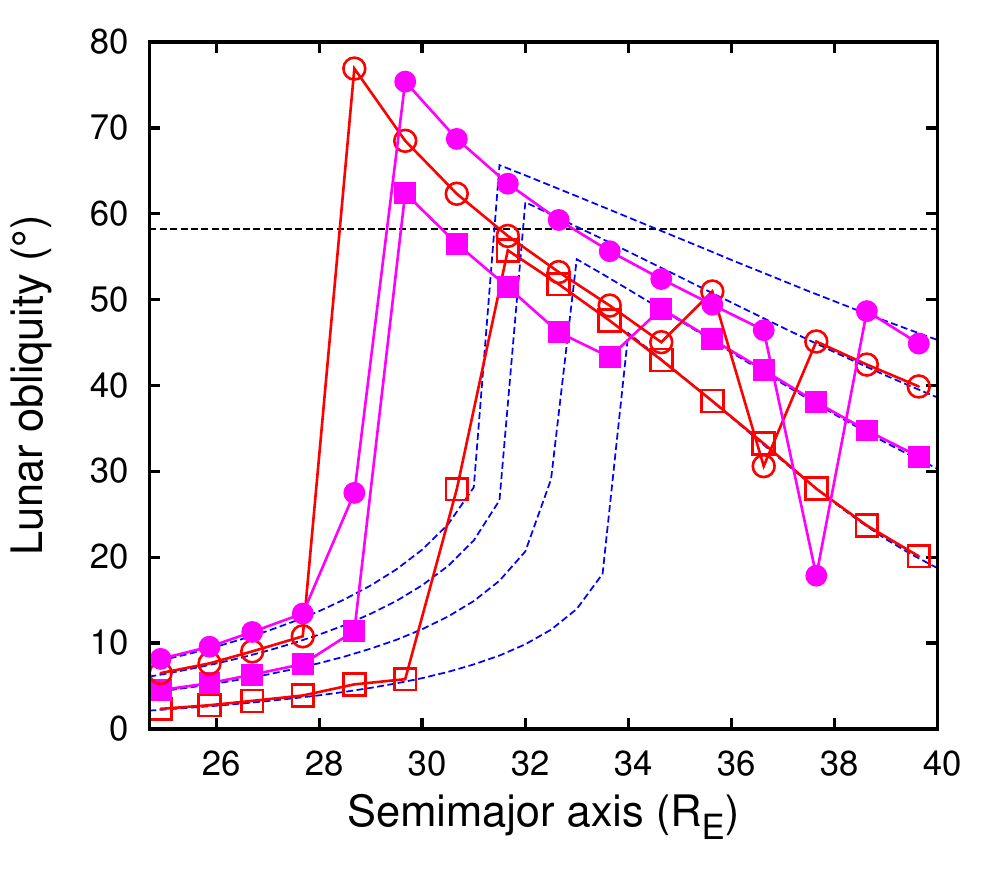}
\caption{\textbf{\textsf{Lunar obliquity close to the Cassini state transition.}} Obliquities for four ``slices" in inclination (at $5^{\circ}$, $10^{\circ}$, $15^{\circ}$, and $20^{\circ}$) from the grid of short simulations shown in Fig. \ref{grid} (solid red and magenta lines with points; the obliquities and inclinations are in the same order at far left and far right). When two different simulations for the same $a$ and $i$ differed in outcome, we chose the solution within the Cassini state, if available. The blue dashed lines plot the relevant Cassini states calculated using analytical formulae, while the black dashed line at $58.15^{\circ}$ plots the upper limit for stable obliquities in the relevant Cassini state. While the numerical and analytical results agree at the smallest and largest semimajor axes, large discrepancies in between are due to non-synchronous rotations being dominant at the Cassini state transition.\label{obl}}
\end{figure}

\begin{figure}
\includegraphics{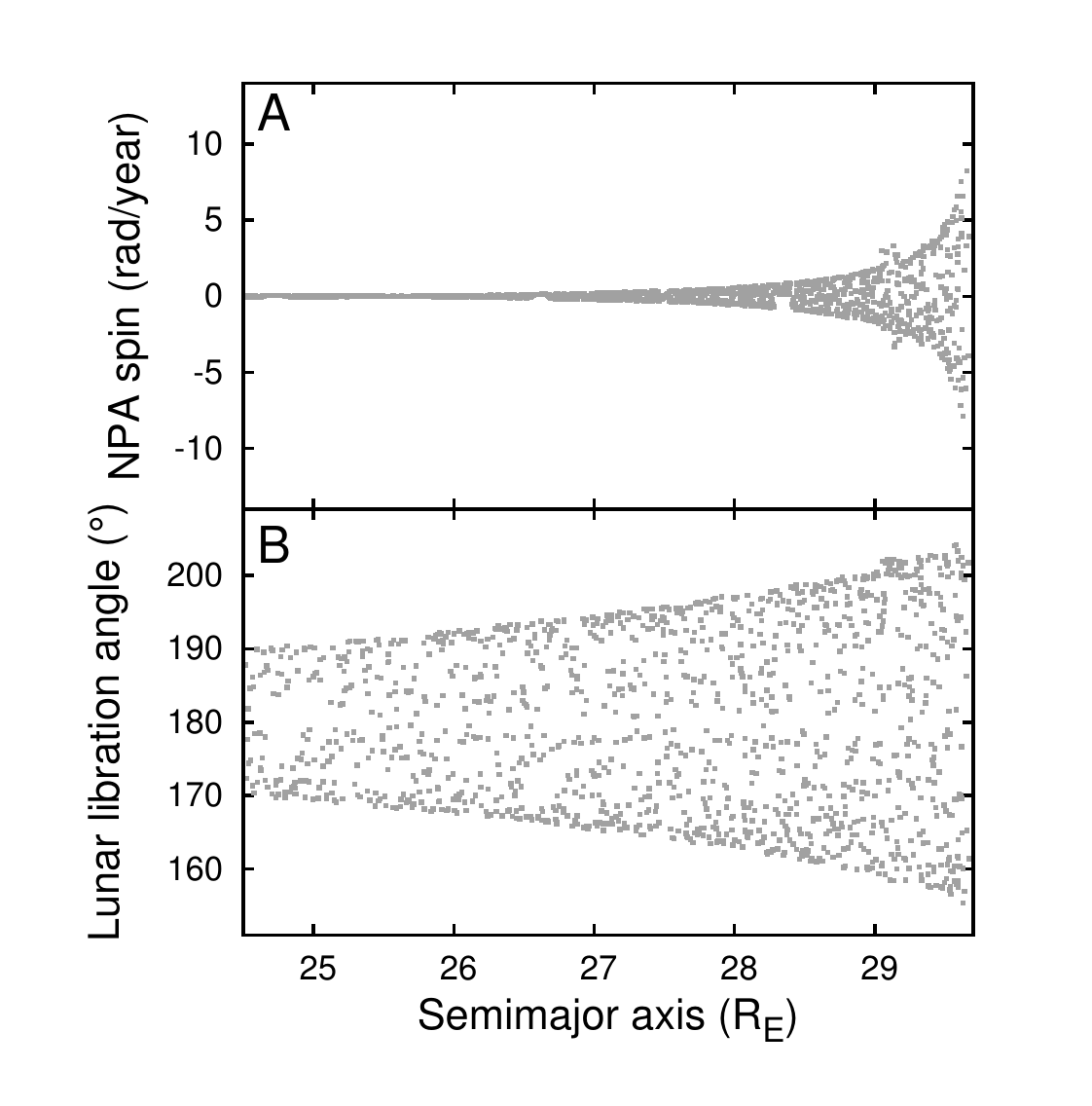}
\caption{\textbf{\textsf{Moon's wobble as it approaches the annual resonance in Fig 4.}} The rotation rate around the longest axis of the Moon (top) and the angle between the longest axis and Earth (bottom) during the first phase of lunar tidal evolution (red points within $29.7 R_E$) in Fig. 4, where we accelerated the tidal evolution by a factor of a hundred. The wobble is clearly building up as the Moon is approaching the resonance between its free wobble and Earth's orbital period at about $29.7 R_E$. The growth in lunar libration angle is more influenced by increasing lunar obliquity (Fig. 4 panel B) than the increase in the amplitude of the wobble.\label{wobble}}
\end{figure}

\begin{figure}
\includegraphics{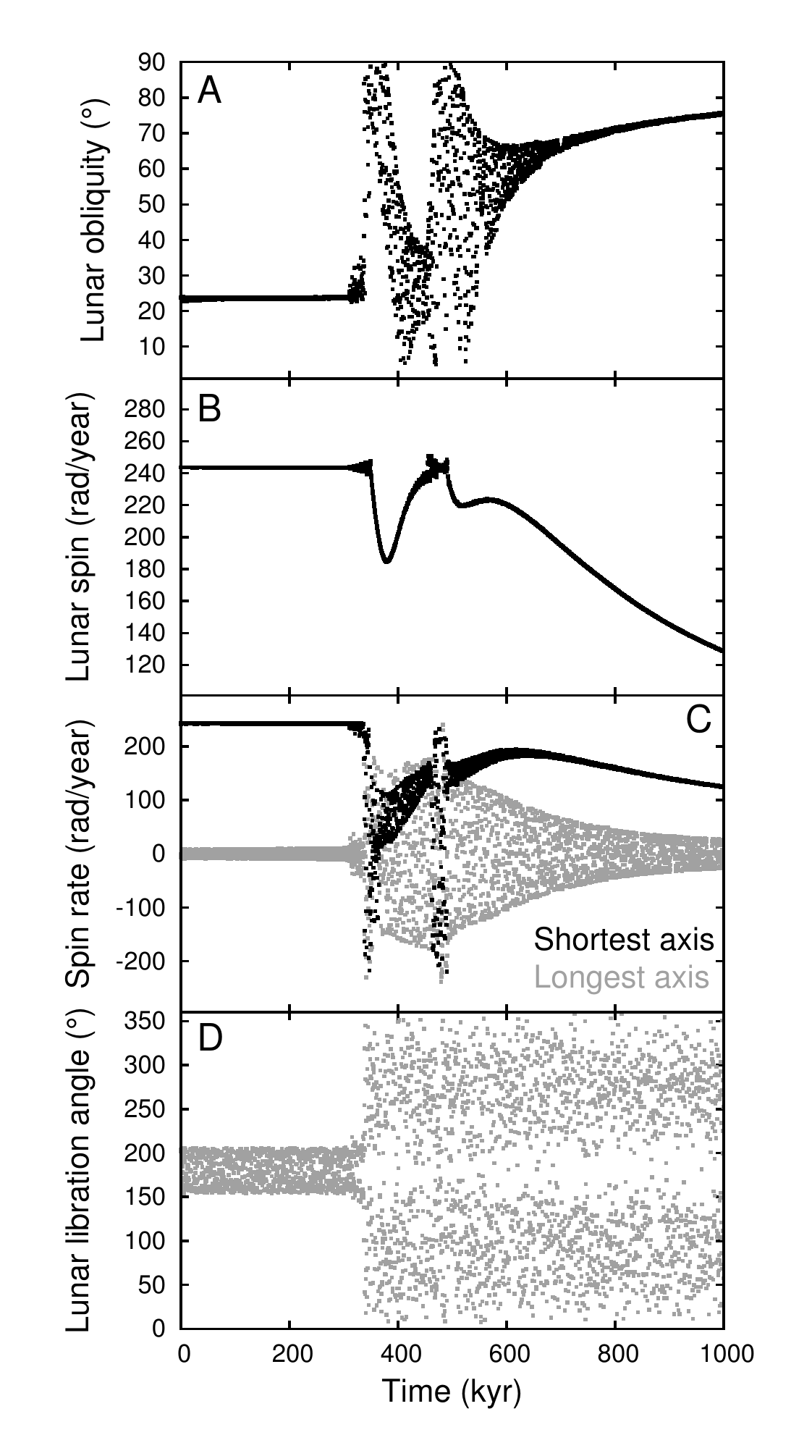}
\end{figure}

\setcounter{figure}{9}

\begin{figure}
\caption{(Previous page) \textbf{\textsf{Passage through the annual resonance of the lunar free wobble in Figure 4.}} Lunar obliquity (A), spin rate (B), rotation rates around the longest and shortest principal axes (C), and the angle between the Moon's longest axis and Earth (D) during the first 1 Myr of the ``blue" segment of lunar tidal evolution in Fig. 4 (which was simulated at the nominal rate for tidal evolution). The free wobble (tracked by gray points in panel C) experiences a resonance at about 330 kyr, breaking the Moon's synchronous rotation. Since the Moon is close to the Cassini state transition, it cannot evolve back into the Cassini state 1 and it settles into a non-synchronous high-obliquity state.\label{res}}
\end{figure}
\thispagestyle{empty}

\end{document}